\documentclass[10pt,aps,prd,showpacs,nofootinbib,twocolumn]{revtex4-1}

\usepackage{amssymb,amsmath,amsfonts,graphicx}
\usepackage{subfig,multirow,txfonts}
\usepackage[usenames]{color}
\usepackage[percent]{overpic}
\usepackage[english]{babel}
\usepackage[bookmarks,colorlinks]{hyperref}
\usepackage[dvipsnames]{xcolor}

\begin{document}

\title{Nonlinear dynamics of oscillating neutron stars in scalar-tensor gravity}

\author{Raissa F.\ P.\ Mendes}
\email{rfpmendes@id.uff.br}
\affiliation{Instituto de F\'isica, Universidade Federal Fluminense (UFF), Niter\'oi, Rio de Janeiro, 24210-346, Brazil.}

\author{N\'estor Ortiz}
\email{nestor.ortiz@nucleares.unam.mx}
\affiliation{Instituto de Ciencias Nucleares, Universidad Nacional Aut\'onoma de M\'exico, Circuito Exterior C.U., A.P. 70-543, M\'exico D.F. 04510, M\'exico.}

\author{Nikolaos Stergioulas}
\email{niksterg@auth.gr}
\affiliation{Department of Physics, Aristotle University of Thessaloniki, Thessaloniki 54124, Greece.}

\date{\today}

\begin{abstract}
The spectrum of oscillating compact objects can be considerably altered in alternative theories of gravity. In particular, it may be enriched by modes with no  counterpart in general relativity, tied to the dynamics of additional degrees of freedom generically present in these theories.
Detection of these modes, e.g.~in the gravitational-wave signal from a binary compact object coalescence, could provide a powerful tool to probe the underlying theory of gravity.
To access the potential of such a detection, it is crucial to understand the linear and nonlinear spectral features of dynamically formed, oscillating compact objects in alternative theories of gravity.
As a step towards that goal, in this work we present a suite of $1+1$ numerical relativity simulations of neutron stars in scalar-tensor theories, we carefully analyze the spectrum of stellar pulsations, and we compare results with expectations from linear perturbation theory. This allows us to build intuition for the case of binary neutron star mergers. Additionally, the models investigated in this work are representatives of two broad classes, in which the scalar field couples either strongly or weakly with the fluid. The distinct phenomenology of the nonlinear dynamics that we identify for each class of models, may find counterparts also in other alternative theories of gravity. 
\end{abstract}


\maketitle

\section{Introduction}

\subsection{Gravitational-wave asteroseismology and alternative theories of gravity}

The operation of the Advanced LIGO detector \cite{advanced-ligo} has allowed the first detection of the fundamental quadrupolar mode of a black hole (BH) in event GW150914~\cite{Abbott2016a}, providing complementary information about the source properties~\cite{LIGOCollaboration2016}. Observation of higher overtones would allow for a direct probe of a BH Kerr nature~\cite{Isi2019,Berti2018}.
Several events that are included in the second gravitational-wave transient catalog (GWTC-2) \cite{Abbott:2020niy}, detected jointly by the Advanced LIGO and Advanced Virgo \cite{advanced-virgo} detectors, are consistent with the presence of an overtone, but these were not yet loud enough to show evidence of deviations from General Relativity (GR) \cite{Abbott:2020jks}. 

Similarly to black hole GW spectroscopy, the observation of the characteristic GW spectrum, or GW asteroseismology, has a rich scientific potential also for neutron stars (NSs) \cite{1999LRR.....2....2K}. 
So far, two GW events have been identified as binary neutron star (BNS) mergers, GW170817 \citep{2017PhRvL.119p1101A} and GW190425 \citep{2020ApJ...892L...3A}, and more are expected in the next years \citep{Aasi:2013wya}. Detection of GWs from the inspiral phase of GW170817, in combination with observations of its electromagnetic counterpart \citep{GBM:2017lvd,Monitor:2017mdv,Goldstein_etal_2017}, have produced new constraints on the dimensionless tidal deformability of NSs and thus on their equation of state (EOS); see \citep{Bauswein_etal_2017,Abbott:2018exr,2020NatAs...4..625C,2020PhRvD.101l3007L,2020Sci...370.1450D,Breschi:2021tbm} 
and references therein, as well as \citep{2020GReGr..52..109C,2021GReGr..53...27D} for recent reviews. These EOS constraints are expected to improve by combining a larger number of detections in the near future \cite{DelPozzo2013,2015PhRvD..92j4008C,2015PhRvD..91d3002L,2019PhRvD.100j3009H,2020PhRvD.101d4019C}.

The frequency of the fundamental $l=2$ quadrupolar mode of nonrotating stars could also be determined from GWs in the inspiral phase of BNS mergers. First constraints (based on GW170817) and future prospects with 3rd-generation detectors were discussed in \citep{2020NatCo..11.2553P}. 
The observation of GWs in the post-merger phase of a BNS merger would offer another opportunity to probe the high-density EOS \citep{1992ApJ...401..226R,PhysRevD.50.6247,PhysRevD.65.103005,2005PhRvL..94t1101S,PhysRevD.81.024012,PhysRevD.83.124008,Stergioulas2011,Bauswein2012,Bauswein2012a,PhysRevD.88.044026,PhysRevD.90.023002,Bauswein2015,PhysRevD.91.064001,Rezzolla2016,2016PhRvD..93f4047D,PhysRevD.101.084039}. Although the sensitivity of the Advanced LIGO and Advanced Virgo detectors was not sufficient to detect the post-merger phase in GW170817  \citep{2017PhRvL.119p1101A,2017ApJ...851L..16A}, such a detection is likely to be achieved in the future, either by upgrading the existing detectors, or with next-generation detectors, see e.g. \citep{2014PhRvD..90f2004C, 2016CQGra..33h5003C, 2017PhRvD..96l4035C, 2018PhRvL.120c1102B, 2018PhRvD..97b4049Y, 2019PhRvD..99d4014T, 2019PhRvD..99j2004M, 2019PhRvD.100d3005E, 2019PhRvD.100d4047T, 2019PhRvD.100j4029B, 2020PhRvD.102d3011E, 2020PASA...37...47A, 2020PhRvL.125z1101H, 2021PhRvD.103b2002G,2021CmPhy...4...27P}.

The remnant formed as a result of a BNS merger has large-amplitude oscillations, with the main peak in the post-merger GW spectrum corresponding to the excitation of the fundamental quadrupolar ($m=2$) $f-$mode. 
The frequency of this main post-merger peak correlates tightly with the radius (and tidal deformability) of nonrotating neutron stars (NSs), leading to improved EOS constrains ~\cite{Bauswein2012,Bauswein2012a,2016EPJA...52...56B,PhysRevD.101.084039}. In addition to the main peak, numerical relativity simulations indicate that nonlinear features may also be present in the GW spectrum of the remnant, due to non-linear combination tones between the $f$-mode and the fundamental quasi-radial ($F$) mode \cite{Stergioulas2011} and due to a transient spiral deformation induced by tidal effects \cite{Bauswein2015}. 
Depending on the EOS, the secondary peaks could become detectable with a detector sensitivity as low as 3.5 times the Advanced LIGO design sensitivity \citep{2019PhRvD..99d4014T}. For more detailed information on the post-merger GW spectrum of BNS mergers, see the reviews \citep{2016EPJA...52...56B,Paschalidis2017,2019JPhG...46k3002B,2020IJMPD..2941015F}.

GWs emitted by oscillating compact objects may also serve as probes of the theory of gravity. Modifications to GR typically imply changes to both equilibrium and perturbative properties of BHs and NSs. Interestingly, even if their equilibrium properties are similar or even identical to their GR counterparts, compact objects in alternative theories of gravity may still display a distinct oscillation pattern, which would imprint characteristic signatures on their GW spectrum (see e.g.~\cite{Barausse2008,Tattersall2018} for examples involving BHs). 

In GR, as a result of reliable numerical relativity simulations and a thorough understanding of perturbation theory, a reasonably clear picture has emerged of the spectral properties of the post-merger GW signal. However, such a clear picture is still missing in many relevant alternative theories of gravity. A primary obstacle to numerical simulations in many of these theories is the lack of a well-posed initial value formulation of their field equations (see e.g.~\cite{PhysRevD.90.124035,Delsate2015,Papallo2017,Cayuso2017,Bernard:2019fjb,Kovacs:2020pns}). Even so, numerical relativity simulations of binary BH mergers have been performed in theories subject to these issues, such as dynamical Chern-Simons~\cite{Okounkova2017,Okounkova2019} or scalar-Gauss-Bonnet gravity~\cite{Witek2019,PhysRevD.102.084046,Silva:2020omi,East:2020hgw,East:2021bqk}, giving hints of the spectral properties of the final BH. However, even for theories with a well-posed initial value formulation, we still lack systematic investigations of the ringdown GW emission, possibly due to the large variety of different models, combined with the numerical cost of performing fully nonlinear $3+1$ simulations.

On the side of perturbation theory, although much has already been done (see e.g.~\cite{Blazquez-Salcedo2019} and references therein), challenges still exist, since many of the useful methods built over the years for GR no longer apply to alternative theories of gravity, thus new techniques are needed (see \cite{langlois2021asymptotics,langlois2021black} for a recent development).
A particularly interesting feature revealed by perturbation theory regards the existence of families of modes with no counterpart in GR, related to the dynamics of the new---scalar, vector, etc.---degrees of freedom generically introduced in alternative theories of gravity. Examples include the presence of both gravitational-led (or fluid-led) and scalar-led modes in scalar-tensor theories of various kinds~\cite{Molina2010,Blazquez-Salcedo2016a,Mendes2018,Blazquez-Salcedo:2020ibb}. When decoupled from fluid/metric perturbations, scalar-led modes describe free oscillations of the scalar field; otherwise, they enrich the combined oscillation spectrum. Since their characteristic frequencies may differ significantly from GR, such scalar-led modes may be promising probes of the underlying theory of gravity. However, to understand how the different modes predicted by perturbation theory are excited in astrophysical situations, such as mergers of BHs and NSs, and access their detectability with next-generation GW detectors, we fall back on the need of nonlinear numerical simulations. This work aims to contribute in this direction.

\subsection{The case of scalar-tensor theories}

To make the discussion more concrete, here we focus on scalar-tensor extensions of GR with action given by
\begin{align} \label{eq:generalaction}
S & = \frac{1}{16 \pi} \int d^4 x \sqrt{-g}
\left[ {\cal R} - 2 \nabla_{\mu} \phi \nabla^{\mu} \phi - V(\phi) \right]
\nonumber \\
& + S_\text{m}[\Psi_\text{m} ; a(\phi)^2 g_{\mu\nu}].
\end{align}
We use geometrized units, such that $c=G= 1$, and denote by $\Psi_\text{m}$ the collection of matter fields, with action $S_\text{m}$. By suitably choosing the model functions $a(\phi)$ and $V(\phi)$, interesting theories can be cast in this format, with diverse phenomenological applications~\cite{Damour1992,Faraoni2004}.
Importantly, scalar-tensor theories (STTs) described by Eq.~(\ref{eq:generalaction}) are known to possess a well-posed initial value formulation~\cite{Salgado2006}, which is crucial for the success of numerical relativity simulations~\cite{Novak1998,Novak2000,Alcubierre2010,Barausse2013, Palenzuela2014,Mendes2016,Gerosa2016,Sagunski2018a,Dima2021}.

Under somewhat general conditions, no-hair theorems guarantee that isolated black holes in theories described by Eq.~(\ref{eq:generalaction}) are those of GR, with a trivial (constant) scalar field profile~\cite{Sotiriou2012}. As a consequence, gravitational and scalar perturbations of BHs will typically decouple at the linear level, and one would not expect scalar perturbations to be significantly excited by a gravitationally perturbed BH (coming, for instance, from a binary BH merger).

On the other hand, NSs are known to exhibit a rich phenomenology in STTs described by Eq.~(\ref{eq:generalaction}). Notably, for appropriate choices of the conformal coupling $a(\phi)$, NSs are prone to \textit{spontaneous scalarization}, a sudden activation of the scalar field as soon as the star reaches a critical compactness~\cite{Damour1993} (see also~\cite{Berti2015} and references therein). This mechanism exemplifies how a theory that resembles GR in the weak field limit can still display a diverse phenomenology in strong-field environments.

The investigation of NS oscillations in STTs started with works by Sotani and Kokkotas~\cite{Sotani2004,Sotani2005}, who derived the full set of perturbation equations around static NSs [with $V(\phi) = 0$], and presented results for the polar and axial sectors. Later developments include more detailed studies of radial oscillations~\cite{Sotani2014}, of EOS effects on the modified spectrum~\cite{AltahaMotahar2018}, of torsional oscillations~\cite{Silva2014}, oscillations of rapidly rotating NSs~\cite{Yazadjiev2017}, and studies in ``massive'' STTs [i.e.~with $V(\phi) \neq 0$], which include the case of $f(R)$ gravity~\cite{Staykov2015,Blazquez-Salcedo2018,Blazquez-Salcedo2019}. 

Remarkably, while axial perturbations have been considered in generality, works on the polar sector (including~\cite{Sotani2004,Sotani2014,Silva2014,Yazadjiev2017,Staykov2015}) have typically resorted to some version of the relativistic Cowling approximation. The primary reason is a technical one: The axial spectrum is affected by the scalar field only through its background value, while in the polar sector scalar and fluid/spacetime perturbations are coupled, making a full treatment more challenging. The situation is considerably simplified if one neglects spacetime oscillations---the Cowling approximation; however, this is often unjustified and may lead to significant errors (see discussion in Ref.~\cite{Mendes2018}; Supplemental Material).
As far as we know, the first study of polar oscillations without the Cowling approximation was done in Ref.~\cite{Mendes2018} by two of the authors, where radial perturbations were analyzed. The full treatment of the problem revealed the existence of a low frequency scalar-led mode, the $\phi$-mode, which is not present within the Cowling scheme. Recently, similar $\phi$-modes have also been reported in tensor-multiscalar theories~\cite{Falcone:2020qqo} and massive Brans-Dicke theories~\cite{Blazquez-Salcedo:2020ibb}. The fundamental $l=2$ mode has also been analyzed without resorting to the Cowling approximation in Refs.~\cite{Blazquez-Salcedo:2020ibb,Kruger:2021yay}.

\begin{figure}[tbh]
\includegraphics[width=0.6\linewidth]{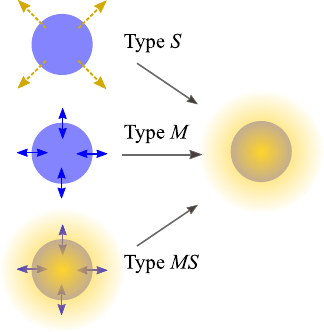}
\caption{
The simulations presented in this work pertain to three types. In all cases, the final state approaches an equilibrium solution supporting a nontrivial scalar cloud. The initial state is close to an equilibrium solution which is either unstable under scalar field (type $S$) or fluid (types $M$ and $MS$) perturbations, and either has a negligible scalar field content (types $S$ and $M$) or already supports a scalar cloud (type $MS$). The labeling stands for spontaneous scalarization ($S$), migration ($M$), and migration from scalarized ($MS$).}
\label{fig:types}
\end{figure}

\subsection{This work}

In order to build intuition about the spectral features of a dynamically formed NS in STTs, in this work we present a suite of six 1$+$1 numerical relativity simulations describing nonlinear radial oscillations of such NSs. We make a thorough analysis of their dynamical evolution, comparing with expectations from linear perturbation theory. These 1$+$1 simulations reveal various ways in which the scalar field can influence the overall evolution, and enable a clean exploration of the excitation and interaction between various modes.

Specifically, we study the excitation of fluid-led and scalar-led modes in the dynamical formation of a \textit{scalarized} neutron star, i.e., a stellar configuration supporting a nontrivial scalar cloud. For that purpose, we investigate three formation scenarios, described pictorially in Fig.~\ref{fig:types}. For the first scenario (type $S$), the initial configuration is an equilibrium solution with negligible scalar field content, but which is unstable under scalar field perturbations. For the other two scenarios, the initial configuration is close to an equilibrium solution that is unstable under fluid perturbations; the initial configuration may be scalarized (type $MS$) or not (type $M$). 

Within each of these scenarios we perform two simulations, corresponding to different choices of the conformal coupling $a(\phi)$ in Eq.~(\ref{eq:generalaction}): The quadratic model of Damour and Esposito-Far\`ese~\cite{Damour1993} with $\beta = -6$ (DEF-6 model) 
and the hyperbolic tangent model introduced by Mendes and Ortiz~\cite{Mendes2016} with $\beta =100$ (MO100 model). 
These models can be seen as representing two broad classes, in which the scalar field is either strongly or weakly coupled to the fluid, respectively. The different kinds of behavior seen in our simulations allow us to extrapolate and identify general features that may also be present in more complex, 3$+$1 simulations, and in other alternative theories of gravity. 

Our work is organized as follows. In Sec.~\ref{sec:setup}, we provide the basic equations relevant for our work, referring the reader to existing literature for details. In Sec.~\ref{sec:linear}, we review equilibrium and perturbative properties of NSs in the scalar-tensor models investigated in this work, laying ground for the analysis of our set of numerical simulations. In Sec.~\ref{sec:nonlinear}, we describe and analyze each simulation in detail. We conclude in Sec.~\ref{sec:implications} with a discussion on the implications of our results for binary NS mergers.

\section{Setup} \label{sec:setup}

\subsection{Theoretical setup} 

\subsubsection{Field equations} \label{sec:setup_theory}
In this work, we restrict attention to the class of massless STTs, which corresponds to $V(\phi)=0$ in action~(\ref{eq:generalaction}). In that case, variation of the action with respect to the metric tensor and scalar field  yields the field equations
\begin{eqnarray}
	G_{\mu\nu} - 2 \nabla_\mu \phi \nabla_\nu \phi + 
	g_{\mu\nu} \nabla_\rho \phi 
	\nabla^\rho \phi  &=& 8\pi a^2 \tilde{T}_{\mu\nu} ,
	\label{eq:G_eq} \\
	\nabla^\mu \nabla_
	\mu \phi &=& -4 \pi a^4 \alpha \tilde{T},
	\label{eq:phi_eq}
\end{eqnarray} 
where 
\begin{equation}
\alpha(\phi) := \frac{d\ln a(\phi)}{d\phi},
\end{equation}
$\tilde{T} := \tilde{g}_{\mu \nu }\tilde{T}^{\mu \nu}$, $\tilde{T}^{\mu\nu} := 2 (-\tilde{g})^{-1/2} \delta S_\text{m} [\Psi_\text{m} ; \tilde{g}_{\rho\sigma} ]/\delta \tilde{g}_{\mu\nu}$ is the (covariantly conserved) Jordan-frame stress-energy-momentum tensor, and (the Jordan-frame metric) $\tilde{g}_{\mu\nu}= a(\phi)^2 g_{\mu\nu}$ is the conformal transformation of the (Einstein-frame) metric $g_{\mu\nu}$.
In this work, we employ the Einstein-frame formulation of STTs, but opt to evolve Jordan-frame fluid variables, since they allow for a more natural interpretation. Quantities associated to the Jordan frame are distinguished with a tilde. 

We model neutron stars by a perfect fluid with energy-momentum tensor
\begin{equation} \label{eq:Tmunu}
\tilde{T}^{\mu\nu} = \tilde{\epsilon} \tilde{u}^\mu \tilde{u}^\nu + \tilde{p} \left(\tilde{g}^{\mu\nu} + \tilde{u}^\mu \tilde{u}^\nu \right),
\end{equation}
where $\tilde{u}^\mu$ is the four-velocity of fluid elements, normalized according to $\tilde{g}_{\mu \nu} \tilde{u}^\mu \tilde{u}^\nu = -1$, and
$\tilde{\epsilon}$ and $\tilde{p}$, respectively, denote the total energy density and pressure of the fluid with respect to comoving observers.

In this work, we shall restrict consideration to spherically symmetric spacetimes, in which case the (Einstein-frame) line element can be written as
\begin{equation}\label{eq:metric1}
ds^2 = - N(t,r)^2 dt^2 + A(t,r)^2 dr^2 + r^2 \left(d\vartheta^2 + \sin^2\vartheta d\varphi^2\right).
\end{equation}
A mass aspect function $m$ can be defined such that $A(t,r)=[1-2m(t,r)/r]^{-1/2}$.

\subsubsection{Equilibrium equations in spherical symmetry}
\label{sec:equilibrium_eqns}

The static, spherically symmetric reduction of the field equations~(\ref{eq:G_eq})-(\ref{eq:phi_eq}), sourced by a perfect fluid with energy-momentum tensor (\ref{eq:Tmunu}), gives rise to the following set of structure equations:
\begin{align}
 &\frac{d m}{dr} = 4\pi r^2 a^4 \tilde{\epsilon} + \frac{r}{2} \left(r-2m\right) \left(\frac{d\phi}{dr}\right)^2, \label{eq:dm}\\
 &\frac{d \ln N}{dr} = \frac{4\pi r^2 a^4 \tilde{p}}{r - 2m} +\frac{r}{2} \left(\frac{d\phi}{dr}\right)^2 + \frac{m}{r(r-2m)}, \label{eq:dn} \\
 &\frac{d^2\phi}{dr^2} = \frac{4\pi r a^4}{r-2m} \! \left[ \alpha (\tilde{\epsilon} - 3\tilde{p}) + r (\tilde{\epsilon} - \tilde{p}) \frac{d\phi}{dr} \right ]\! -\frac{2(r-m)}{r(r-2m)} \frac{d\phi}{dr}, \label{eq:dphi} \\
 &\frac{d\tilde{p}}{dr} = -(\tilde{\epsilon} + \tilde{p}) \left[  \frac{4\pi r^2 a^4 \tilde{p}}{r-2m} \! + \! \frac{r}{2} \left(\frac{d\phi}{dr}\right)^2 \!\! + \! \frac{m}{r(r-2m)} \! + \! \alpha \frac{d\phi}{dr} \right]. \label{eq:dp}
\end{align}
An equation of state is required in order to close the system~(\ref{eq:dm})-(\ref{eq:dp}). In this work, we adopt the polytropic EOS described in Sec.~\ref{sec:EoS}.

Given the central pressure of a star, $\tilde{p}_c$, and an asymptotic value for the scalar field, $\phi_0$, Eqs.~(\ref{eq:dm})-(\ref{eq:dp}) can be integrated by standard methods, subject to suitable regularity conditions (see, e.g.~Sec.~II-B of Ref.~\cite{Mendes2016}).
Each stellar configuration is characterized by its ADM mass $M$, radius $R$, and scalar charge $Q_s$, defined through the asymptotic expansion $\phi = \phi_0 + Q_s/r + O(r^{-2})$.

\subsubsection{Evolution equations in spherical symmetry}

We solve numerically the field equations~(\ref{eq:G_eq})-(\ref{eq:phi_eq}) in spherical symmetry, coupled to the fluid equations
\begin{eqnarray}
\tilde{\nabla}_\nu \tilde{T}^{\mu\nu} &=& 0,\label{eq:T_conserv} \\
\tilde{\nabla}_\mu\tilde{J}^\nu &=& 0,\label{eq:MB_conserv}
\end{eqnarray}
where $\tilde{J}^\mu = \tilde{\rho} \tilde{u}^\mu$ is the baryon mass current, with $\tilde{\rho}$ the rest mass density measured by comoving observers. 
We work in the radial gauge with polar slicing condition~\cite{Alcubierre-Book}, thus the line element in the Einstein frame has the form~(\ref{eq:metric1}).

The Finite Volume method employed in this work---see Sec.~\ref{sec:setup_numerics} for details---requires the evolution equations to be written as a hyperbolic system of conservation laws of the form
\begin{equation}\label{eq:flux-conservative-form}
\frac{\partial}{\partial t}(A {\bf q}) + \frac{1}{r^2} \frac{\partial}{\partial r}\left(NA r^2 {\bf F}({\bf q})\right) = {\bf S}({\bf q}),
\end{equation}
with the vector of conserved quantities ${\bf q} = ( \tilde{D}, \tilde{S}, \tilde{\tau} , \eta, \psi )^\text{T}$, the flux vector
${\bf F} = (F_{\tilde{D}}, F_{\tilde{S}}, F_{\tilde{\tau}}, F_\eta, F_\psi)^\text{T}$, and the source vector ${\bf S} = (S_{\tilde{D}}, S_{\tilde{S}}, S_{\tilde{\tau}}, S_\eta, S_\psi)^\text{T}$.
The baryon mass density $\tilde{D}$, radial momentum density $\tilde{S}$, internal energy density $\tilde{\tau}$, and total energy density $\tilde{E}$, are all conserved quantities measured by Eulerian observers. They are given by
\begin{eqnarray}
\tilde{D} &:=& \tilde{\rho} \Gamma, \label{eq:D} \\
\tilde{S} &:=& (\tilde{E} + \tilde{p}) A^2 v , \label{eq:S}\\
\tilde{\tau} &:=& \tilde{E} - \tilde{D}, \label{eq:tau}\\
\tilde{E} &:=& \Gamma^2 (\tilde{\epsilon} + \tilde{p}) - \tilde{p}, \label{eq:E}
\end{eqnarray}
where the Lorentz factor can be written as
\begin{equation}
\Gamma = \left(1 - A^2 v^2 \right)^{-1/2},
\end{equation}
with $A v$ the fluid's radial velocity measured by Eulerian observers.
The scalar field variables $\eta$ and $\psi$ are defined by
\begin{equation}\label{eq:eta_and_psi}
\eta := \frac{1}{A} \frac{\partial \phi}{\partial r}, \qquad
\psi := \frac{1}{N} \frac{\partial \phi}{\partial t}.
\end{equation}
Explicit expressions for the flux and source vectors involved in Eq.~(\ref{eq:flux-conservative-form}) can be found in Sec.~II-A of Ref.~\cite{Mendes2016}.

We evolve the mass aspect function $m(t,r)$ through
\begin{equation}
\frac{\partial m}{\partial t} = r^2 \frac{N}{A^2} \left( A \eta\psi - 4\pi a^4 \tilde{S} \right),\label{eq:momentum_constraint}
\end{equation}
and we solve for the lapse function $N(t,r)$ at each time slice using
\begin{equation}
\frac{\partial N}{\partial r} = A^2 N \left[ \frac{m}{r^2} + 4\pi r a^4 \left( \tilde{p} + \tilde{S} v \right) + \frac{r}{2} \left( \eta^2+\psi^2 \right) \right].\label{eq:lapse_condition}
\end{equation}

\subsubsection{Radial perturbation equations}
\label{sec:radialeqs}

In order to interpret results from our nonlinear simulations, it is important to understand the radial spectrum predicted by linear perturbation theory. Thus, we consider linear radial perturbations of the equilibrium stellar configurations described in Sec.~\ref{sec:equilibrium_eqns}. 
We represent scalar field and metric radial perturbations by
\begin{eqnarray}
\phi(t,r) &=& \phi_{(0)}(r) + \delta\phi(t,r),\\
N(t,r) &=& N_{(0)}(r) + \delta N (t,r),\\
A(t,r) &=& A_{(0)}(r) + \delta A(t,r),
\end{eqnarray}
where the subscript $(0)$ refers to background quantities. Correspondingly, the Jordan-frame metric perturbation can be written as $\tilde{g}_{\mu\nu} = \tilde{g}_{\mu\nu}^{(0)} + \tilde{h}_{\mu\nu}$, where $\tilde{h}_{\mu\nu} = a_{(0)}^2 h_{\mu\nu} + 2 g_{\mu\nu}^{(0)}  a_{(0)}^2 \alpha_{(0)} \delta \phi$, $a_{(0)} := a(\phi_{(0)})$, and $\alpha_{(0)} := \alpha(\phi_{(0)})$.

Regarding the perturbed fluid, it is assumed to be described by the same cold EOS as the background configuration, and its perturbation is characterized by the Lagrangian displacement vector $\tilde{\xi}^\mu=(0, \xi, 0, 0)^\text{T}$. 
Pressure and energy density perturbations, $\delta \tilde{p}$ and $\delta \tilde{\epsilon}$, are given by $\delta \tilde{p}  = \tilde{\rho}^{-1} \Gamma_1 \tilde{p} \delta \tilde{\rho}$ and $\delta \tilde{\epsilon} = \tilde{\rho}^{-1}(\tilde{\epsilon} + \tilde{p}) \delta \tilde{\rho}$, with $\delta\tilde{\rho}$ being the rest-mass perturbation, and $\Gamma_1 := \partial \ln \tilde{p}/ \partial \ln \tilde{\rho}$ the adiabatic exponent.

All perturbation variables can be written solely in terms of $\xi$ and $\delta \phi$, which obey a system of coupled, second-order, master equations. In order to obtain the linear spectrum, we search for solutions of the form 
\begin{equation}
\xi(t,r) = \xi (r) e^{i\omega t}, 
\quad
\delta \phi (t,r) = \delta \phi (r) e^{i\omega t},
\quad
\omega \in \mathbb{C},
\end{equation}
and cast the master equations in the form of a homogeneous system of first-order, coupled ordinary differential equations (ODEs),
\begin{equation} \label{eq:perturbation_equation}
    \frac{d {\bf x} (r)}{dr} =  {\bf M}(r) {\bf x}(r),
\end{equation}
where ${\bf x}(r) = (\xi, d\xi/dr, \delta \phi, d \delta \phi/dr)^\text{T}$ and ${\bf M}(r)$ is a 4 x 4 matrix with coefficients that depend only on background quantities. The vector ${\bf x}(r)$ is required to be everywhere regular and $\delta\phi(t,r)$ to be purely outgoing at spatial infinity.

The explicit form of ${\bf M}(r)$ can be found in the Supplemental Material of Ref.~\cite{Mendes2018}, but it is worthwhile to point out that the components that couple fluid and scalar field perturbations are all proportional to either $\alpha_{(0)}$ or $d\phi_{(0)}/dr$. When these quantities vanish, ${\bf M}(r)$ becomes block diagonal, and fluid and scalar field perturbations decouple.

We refer to the Supplemental Material of Ref.~\cite{Mendes2018} for a thorough discussion on the relevant boundary conditions for the radial mode analysis, integration methods, as well as expressions for the remaining perturbation variables $(\delta N, \delta A, \delta\tilde{\rho})$, which can be recovered from $\xi$ and $\delta\phi$.

\subsubsection{Coupling functions}
\label{sec:setup_coupling}

A particular (massless) STT model is defined by the coupling function $a(\phi)$.
Here, we consider two STT models, the DEF (Damour-Esposito-Far\`ese) model~\cite{Damour1993}, which is the simplest one leading to spontaneous scalarization, and the MO (Mendes-Ortiz) model~\cite{Mendes2016}, which corresponds to an analytical approximation to the coupling function of a standard massless scalar field nonminimally coupled to gravity. These models are represented, respectively, by the coupling functions
\begin{align}
\textrm{\bf DEF:}\quad &a(\phi) = e^{\frac{1}{2}\beta \phi^2}, \nonumber \\
& \alpha(\phi) = \beta \phi, \label{eq:DEF} \\
\textrm{\bf MO:}\quad &a(\phi) = \left[ \cosh \left(\sqrt{3} \beta \phi\right) \right]^\frac{1}{3\beta}, \nonumber \\
& \alpha(\phi) = \frac{1}{\sqrt{3}} \tanh [\sqrt{3} \beta \phi]; \label{eq:MO}
\end{align}
where $\beta$ is a real parameter determining the leading order behavior of both models around $\phi \approx 0$. 

While the DEF model only allows for spontaneous scalarization when $\beta$ is sufficiently negative (specifically $\beta \lesssim -4.35$ ~\cite{Harada1998,Palenzuela2016}), spontaneous scalarization also occurs in the MO model for sufficiently large, positive values of $\beta$, as long as the NS EOS is sufficiently stiff \cite{Mendes2016}. In the following, we will consider the DEF model with $\beta = -6$ (DEF-6) and the MO model with $\beta = 100$ (MO100). 

Further details on the motivation and properties of the coupling functions~(\ref{eq:DEF}) and~(\ref{eq:MO}) can be found in Sec.~III-B of Ref.~\cite{Mendes2016}.

\subsubsection{Equation of state}
\label{sec:EoS}

In this work, we assume a polytropic EOS,
\begin{equation}
	\tilde{p}(\tilde{\rho}) = K \rho_0 (\tilde{\rho}/\rho_0)^\gamma,
\end{equation}
with $\gamma = 3$, $K = 0.005$, and $\rho_0 = 1.66 \times 10^{14}$g/cm${}^3$. Evidently, our EOS renders rather simplistic and not fully realistic NS models. Although our EOS can mimic realistic NS core matter~\cite{Read2009} and produce a mass-radius relation consistent with observational constraints~\cite{Ozel2016}, it suffers from superluminal sound speeds at sufficiently large densities, as well as a poor description of NS crust matter. These issues can be alleviated by small modifications of the EOS parameters, which would not alter our results, at least qualitatively.

More importantly, our choice of EOS guarantees the existence of hydrodynamically stable NSs with compactness $M/R \gtrsim 0.27$, so that the trace of the energy-momentum tensor, $\tilde{T} = 3\tilde{p} - \tilde{\epsilon}$, is positive near the center of the star, which is a necessary condition to allow for spontaneous scalarization in STT models with $\beta >0$~\cite{Mendes2015}.

\begin{figure*}[t]
\centering
\includegraphics[width=0.85\textwidth]{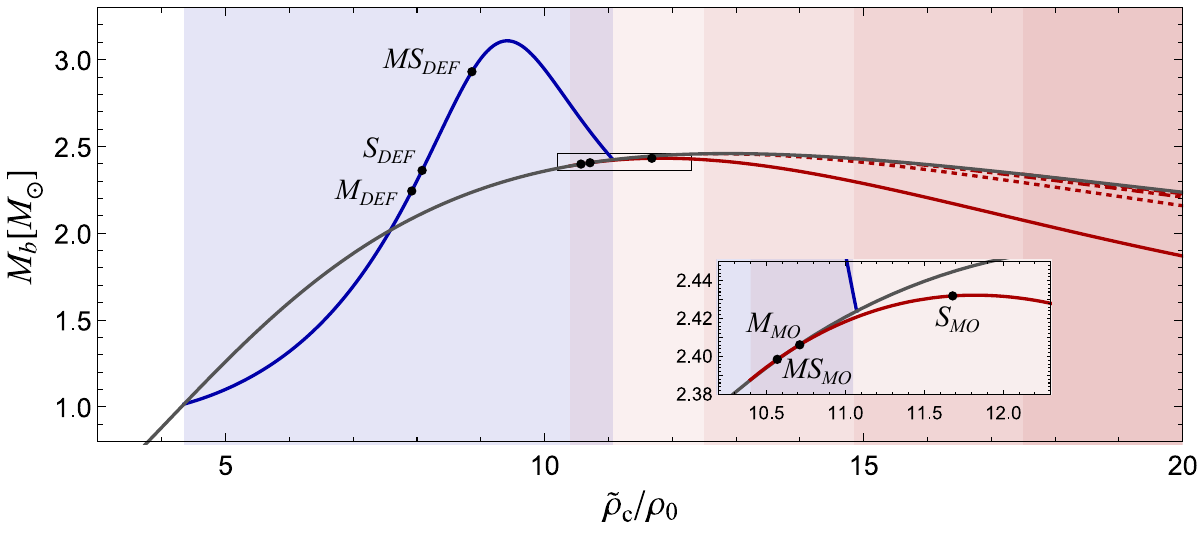}
\caption{Baryon mass as a function of the central rest-mass density for equilibrium solutions in the models studied in this work. The gray curve corresponds to equilibrium configurations shared by GR and STTs. For $4.35 \rho_0 \lesssim \tilde{\rho}_c \lesssim 11.07 \rho_0$ (blue vertical band), the DEF-6 model admits two additional equilibrium solutions, with the same baryon mass but opposite scalar charges. These sequences of scalarized solutions are shown in blue. For $\tilde{\rho}_c \gtrsim 10.38 \rho_0$, the MO100 model also admits two new equilibrium solutions, with the same baryon mass and opposite scalar charges. These sequences correspond to the solid red curve. Moreover, the MO100 model admits additional branches of equilibrium solutions which detach from the GR branch at increasing values of the central rest-mass density (at $\tilde{\rho}_c = 12.46 \rho_0$, $14.80 \rho_0$, $17.47 \rho_0,~\dots$ These intervals are highlighted as vertical red bands of different opacities). Black dots locate the final equilibrium solutions of the six simulations reported in this work (see Table \ref{tab:simulations}). The inset highlights the portion of the diagram relevant for simulations in the MO100 model.}
\label{fig:equilibrium}
\end{figure*}

\subsection{Numerical setup} \label{sec:setup_numerics}

In this Section, we describe the numerical methods and spectral analysis tools employed throughout this work.

\subsubsection{Non-linear simulations}
We solve the hyperbolic system of conservation laws~(\ref{eq:flux-conservative-form}) using a Finite Volume numerical scheme together with a Harten-Lax-van Leer-Einfeldt approximate Riemann solver~\cite{Harten_etal83,Einfeldt88}, and a linear piecewise variable reconstructor~\cite{LeVeque-Book_FV}.
For time evolution, we employ a third-order Runge-Kutta integrator. Both space and time coordinates are discretized in uniform numerical grids, with a typical spatial resolution $\Delta r$ of the order of $10^{-4}r_\textrm{max}$, where $r_\textrm{max}$ is the outer boundary of the radial domain, chosen such that $r_\textrm{max}/R \sim 20$, where $R$ is the surface of the star. Time step-size $\Delta t$ is determined by the Courant-Friedrichs-Lewy condition $\Delta t/\Delta r = \kappa_\textrm{CFL} < 1$, with $\kappa_\textrm{CFL} = 0.25$.

\begin{itemize}
    \item {\it Initial data.}
    Our non-linear time evolutions use initial data consisting of numerical solutions of the ODE system~(\ref{eq:dm})-(\ref{eq:dp}); see the Appendix of Ref.~\cite{Mendes2016} for details.
    
    \item {\it Migration simulations.}
    For dynamical scenarios type $M$ and $MS$, which consist on migration from the unstable branch of equilibrium stellar configurations to the stable branch while keeping roughly the same baryonic mass (see Fig.~\ref{fig:equilibrium} and Sec.~\ref{sec:nonlinear}), we enforce migration by radially perturbing the baryonic mass initial profile, namely $\tilde{\rho} \to \tilde{\rho} (1 - \varepsilon \cos[(\pi/2)r/R])$~\cite{Font2002}. The small perturbation parameter $\varepsilon$ ranges from $0.01$ to $0.05$, and translates into $\Delta\tilde{\rho}_c$ in Table~\ref{tab:simulations}.
    Such a small perturbation is sufficient to dominate over truncation errors that would otherwise lead to gravitational collapse.
    After perturbing the original equilibrium configuration, we enforce the constraints in order to guarantee initial data consistency with the field equations.

    \item {\it Boundary conditions.}
    According to the spherical symmetry of the problem, we demand the quantities $(\tilde{D},\tilde{\tau},\psi)$ to be even functions near $r = 0$ during the whole time evolution, and $(\tilde{S},\eta)$ to be odd functions. At the outer boundary, $r=r_\textrm{max}$, we impose the outgoing flow condition $\left. \partial_r (\tilde{D},\tilde{S},\tilde{\tau}) \right|_{r_\textrm{max}} = 0$ at all times. For the scalar field, the outgoing wave condition $\phi(t,r) \overset{r \to \infty}{\longrightarrow} F(t-r)/r$, where $F$ is an arbitrary function of retarded time, translates into~\cite{Novak1998b}
    \begin{equation}\label{eq:outgoing_condition}
    \left.\left( A\eta + A\psi + \frac{\phi}{r} \right)\right|_{r_\textrm{max}} = 0.
    \end{equation}
    
    \item {\it Vacuum treatment.}
    We employ the standard ``atmosphere'' numerical artifact~\cite{jFmMwSmT00} to consistently solve for the fluid variables in the exterior region of compact stars. This strategy consists in replacing the (ideal) vacuum with a constant baryon density which is several orders of magnitude smaller than the central density of the star. We follow the implementation described in Ref.~\cite{jFmMwSmT00}, with an atmosphere density $\rho_\textrm{atm} \sim 10^{-11}\tilde{\rho}_c$, where $\tilde{\rho}_c$ is a given stellar central density.
    
    \item {\it Convergence.}
    During time evolution, we monitor violations of the Hamiltonian constraint, the scalar-field constraint, and baryonic mass conservation, respectively given by
    \begin{eqnarray}
    {\cal H} &=& \frac{\partial m}{\partial r} - \frac{r^2}{2} \left[ \eta^2 + \psi^2 + 8\pi a^4 \left( \tilde{\tau} + \tilde{D} \right) \right], \label{eq:Hamiltonian} \\
    {\cal C} &=& \partial_r\phi -  A\eta,\\ \label{eq:wave_constrint}
    {\cal M} &=& M_b - \int_0^R 4 \pi r^2 \tilde{D}~a(\phi)^3( 1- 2m/r)^{-1/2} dr,\label{eq:Total_BM}
    \end{eqnarray}
    where $M_b$ is the initial baryon mass. We evaluate the $L_1$-norm of constraint violations at each time step for different resolutions. We observe self-convergence to order $\gtrsim 1.5$ for the Hamiltonian constraint. For the scalar field constraint, we obtain second-order self-convergence.
    Regarding baryonic mass, we have verified a conservation trend as we increase numerical resolution. For production runs, violations of baryonic mass conservation are at most of the order of 1 part in $10^2$. For details on convergence tests, we refer to the Appendix of Ref.~\cite{Mendes2016}.
\end{itemize}

\subsubsection{Fourier analysis}

In order to analyze the spectral features of the simulations, we record the time evolution of some evolved quantities at fixed points in the spatial grid. This gives rise to discrete time series, which are converted to their frequency domain representation by a Discrete Fourier Transform (DFT). Before performing the DFT, irrelevant low-frequency components were removed from the data by subtracting from the time series a quadratic fit to it.

In DFT plots presented in Sec.~\ref{sec:nonlinear}, we include the DFT at various spatial points, as a means to unambiguously identify the presence of the various modes. The mean amplitude of the DFT along the radial direction is also represented. From the recorded data, we can also uncover some features of the wavefunctions associated with each frequency peak, by exploring the correlation between these wavefunctions and the real part of the DFT as a function of the radial coordinate, for a fixed frequency. In particular, this allows us to identify the presence of nodes in the wavefunctions, which is a distinguishing property of the various modes.

\section{Equilibrium and perturbative properties} \label{sec:linear}

Before we turn to the nonlinear simulations and their spectral analysis, in this section we briefly review some relevant equilibrium properties of NSs in STTs, as well as their behavior under radial perturbations. This will provide background to the interpretation of our results in Sec.~\ref{sec:nonlinear}.

Below a certain critical compactness, NSs in STTs are similar to their GR counterparts. Deviations depend on the background value of the scalar field ($\phi_0$) in which they are immersed, and on the model parameters. 
For isolated NSs, $\phi_0$ is of cosmological origin, and it is constrained by solar system experiments to be close to zero for the models addressed in this work \cite{Will1993}.
However, it should be noted that the ambient scalar field may effectively differ from the asymptotic, cosmological value if, e.g., the NS is in a close orbit with a companion supporting a nontrivial scalar cloud (which gives rise, in particular, to the effect of dynamical scalarization \cite{Barausse2013,Palenzuela2014,Sampson2014,Taniguchi2015}). We will refer to that situation occasionally, but for simplicity will set $\phi_0 = 0$ in what follows.

For $\phi_0$ identically zero, low compactness equilibrium solutions are therefore identical to those in GR, and are characterized by a trivial scalar field content ($\phi = \phi_0=0$).  However, above a certain critical compactness, new equilibrium solutions appear, beyond the GR-like one. They exhibit a nontrivial scalar field profile and are known as scalarized configurations. Sequences of equilibrium solutions are shown in Fig.~\ref{fig:equilibrium} for the two models addressed in this work. 
In the DEF-6 model, branches of scalarized solutions exist in the range $4.35 \lesssim \tilde{\rho}_c/\rho_0 \lesssim 11.07$ of central densities, and have a maximum mass larger than that of GR. Along the branches of scalarized solutions, the absolute value of the scalar charge increases from zero to a maximum value of $ |Q_s|/M \approx 0.648$, and again decreases to zero at the boundary of the scalarization region. 
In the MO100 model, branches of scalarized solutions exist for central densities $\tilde{\rho}_c \gtrsim 10.38 \rho_0$, and have a smaller maximum mass than that of GR. Along the branches of scalarized solutions, the absolute value of the scalar charge grows monotonically from zero; however, typical values are much smaller than in the DEF-6 model: For the maximum mass solution, $ |Q_s|/M \approx 1.60 \times 10^{-4}$. The small values of scalar charges found in the MO model with $\beta>0$ are the main reason why it can evade pulsar timing constraints \cite{Mendes2019}, which have provided stringent tests for the $\beta <0$ case \cite{Freire2012, Anderson2019a}. 

\begin{figure}[thb]
\centering
\includegraphics[width=0.45\textwidth]{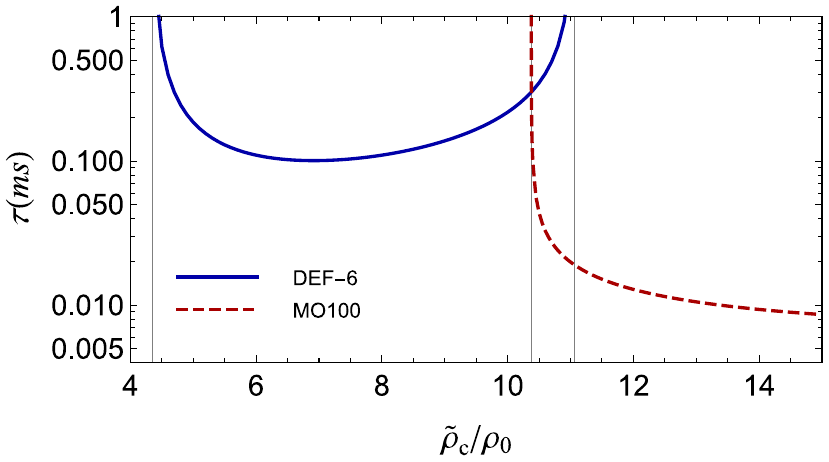}
\caption{Timescale for the exponential growth of scalar perturbations around GR-like solutions for the two scalar-tensor models studied in this work. Vertical lines show the boundaries of the instability region. The instability timescale for the MO100 model is seen to be much shorter than that of the DEF-6 model.}
\label{fig:instability}
\end{figure}

Let us now turn to the radial spectrum of the equilibrium solutions described above (see also Ref.~\cite{Mendes2018}).
For GR-like solutions, with a trivial scalar field content ($\phi = \phi_0 = 0$), radial fluid perturbations decouple from scalar field perturbations (cf.~Sec.~\ref{sec:radialeqs}).
In this case, the fluid spectrum is identical to GR. On the other hand, scalar perturbations around these GR-like solutions become unstable in the range of densities where scalarized equilibrium solutions exist \cite{Harada1997}. The instability timescale is shown in Fig.~\ref{fig:instability} for the two models considered in this work. As we will discuss below, this instability is driven by the fundamental scalar-led mode, which we will also refer to as $\phi$-mode. As the critical point for spontaneous scalarization is approached, the frequency of the $\phi$-mode goes to zero, and it becomes purely imaginary beyond that point if one follows along the sequence of GR-like solutions. 

\begin{figure}[thb]
\centering
\includegraphics[width=0.45\textwidth]{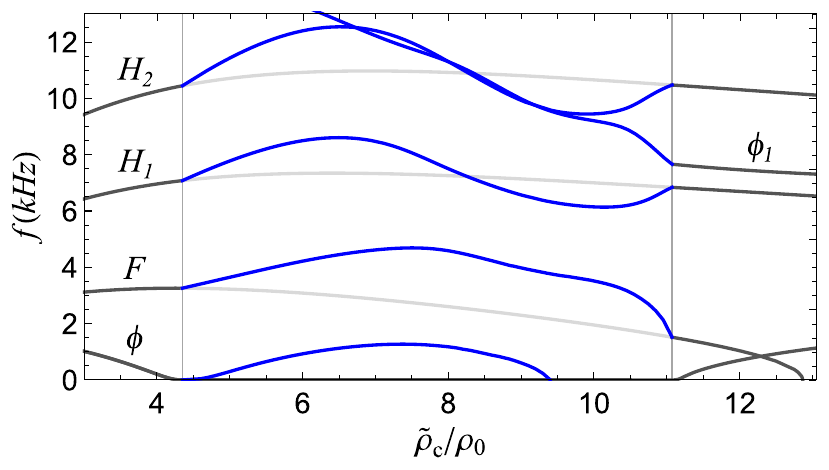}
\caption{Frequency as a function of central rest-mass density for radial modes in the DEF-6 model. Vertical lines delimit the scalarization region. In this range of central densities, GR mode frequencies are shown in light gray for comparison.}
\label{fig:linearspectrumDEF}
\end{figure}

\begin{figure*}[t]
\centering
    \includegraphics[width=0.7\textwidth]{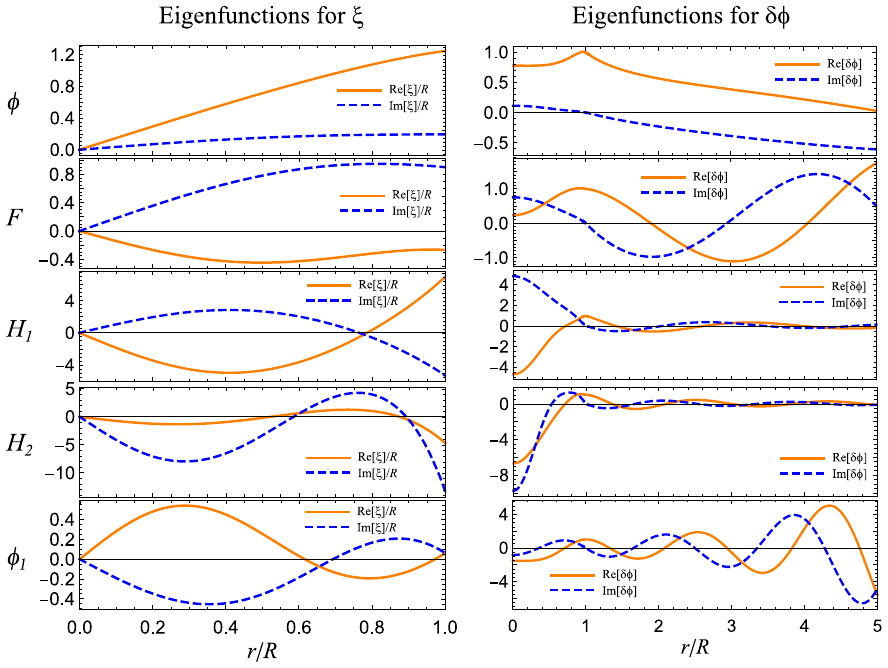}
    \caption{Real (orange) and imaginary (dashed blue) parts of the eigenfunctions for the Lagrangian displacement $\xi$ and scalar field perturbation $\delta \phi$ for the five lower frequency modes of a scalarized NS with $\tilde{\rho}_c = 8.08 \rho_0$ in the DEF-6 model.
    }
    \label{fig:DEF_eigenfunctions}
\end{figure*}

On the other hand, for a background scalarized solution, fluid and scalar field perturbations are coupled. This coupling continuously changes the angular frequency $\omega$ of the GR fluid modes, and in particular introduces an imaginary part to these frequencies which accounts for damping due to scalar radiation. Figure \ref{fig:linearspectrumDEF} shows the frequency $f = \Re(\omega)/(2\pi)$ as a function of the central density for the lowest-frequency modes found in the DEF-6 model. Besides the modified (fluid-led) $F$, $H_1$, and $H_2$ modes, one can identify the presence of two scalar-led modes, namely the fundamental $\phi$-mode and the first overtone $\phi_1$. In particular, an important role is played by the $\phi$-mode. As anticipated above, at the boundaries of the scalarization region, the $\phi$-mode has zero frequency, and it is the lowest frequency (fundamental) mode of the branch of scalarized solutions that emerges at that point. Importantly, it is the $\phi$-mode, not the fluid-led $F$-mode, that governs the stability of scalarized solutions under gravitational collapse in the scalarization region. Indeed, it becomes unstable in the range of central densities $9.41 \lesssim \tilde{\rho}_c/\rho_0 \lesssim 11.07$, which starts at the turning point of the $M_b-\tilde{\rho}_c$ diagram in Fig.~\ref{fig:equilibrium}. The fact that stability under gravitational collapse is governed by a scalar-led mode is a result of the strong coupling between fluid and scalar field perturbations in the DEF-6 model. This coupling depends on derivatives of the background scalar field, and the fact that such derivatives are large for the DEF-6 model (in comparison to MO100) is behind the larger scalar charges displayed by scalarized solutions in this case. Another consequence of this strong coupling is that energy can be efficiently transferred and radiated away by the scalar field. The damping timescale $\tau_\text{damp} = \Im(\omega)^{-1}$ is therefore typically short: $\tau_\text{damp} \gtrsim 0.10$ ms for the $\phi$-mode, $\tau_\text{damp} \gtrsim 0.079$ ms for the $F$-mode, $\tau_\text{damp} \gtrsim 0.60$ ms for the $H_1$ mode, $\tau_\text{damp} \gtrsim 0.63$ ms for the $H_2$ mode, and $\tau_\text{damp} \gtrsim 0.024$ ms for the $\phi_1$ mode.

Figure \ref{fig:DEF_eigenfunctions} illustrates the eigenfunctions for the $\phi$, $F$, $H_1$, $H_2$, and $\phi_1$ modes in model DEF-6, considering the case of a scalarized solution with $\tilde{\rho}_c = 8.08 \rho_0$ (which is the central density of the scalarized NS formed in simulation $S_\text{DEF}$; see below). The Lagrangian displacement has no nodes for the $\phi$ and $F$ modes, for the $H_1$ mode it has one node, and for the $H_2$ mode it has two nodes. The $\phi_1$ mode has a Lagrangian displacement with two nodes in its real part and one node in its imaginary part. Note that the asymptotic behavior of the scalar field perturbation for a mode with frequency $\omega$ is $e^{-i \omega (t - r)}$: The larger $\Im(\omega)$, the stronger is the damping and the more pronounced is the growth of the radial part of the $\delta \phi$-eigenfunction as $r$ grows large; see Fig.~\ref{fig:DEF_eigenfunctions}.

\begin{figure}[thb]
\centering
\includegraphics[width=0.45\textwidth]{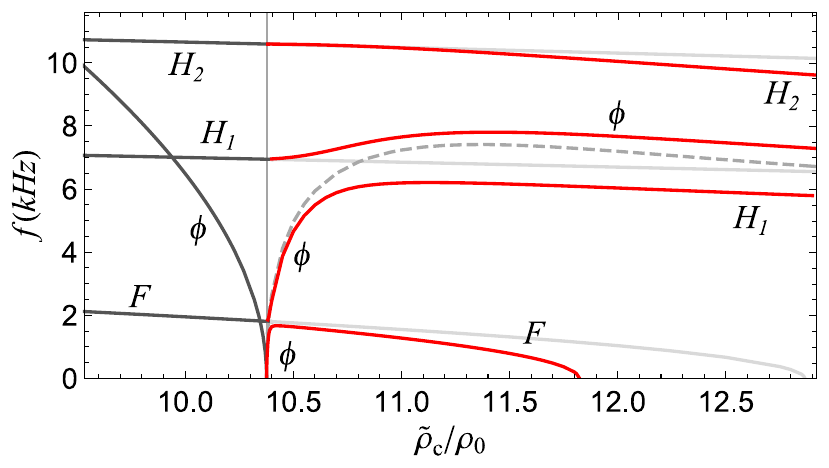}
\caption{Frequency as a function of central rest-mass density for radial modes in the MO100 model. A vertical line delimits the scalarization region. In this range of central densities, GR mode frequencies are shown in light gray for comparison. Additionally, a dashed gray curve shows the frequency of the fundamental scalar mode when fluid perturbations are frozen; see main text for details.}
\label{fig:linearspectrumMO}
\end{figure}

\begin{figure*}[ht]
\centering
    \includegraphics[width=0.7\textwidth]{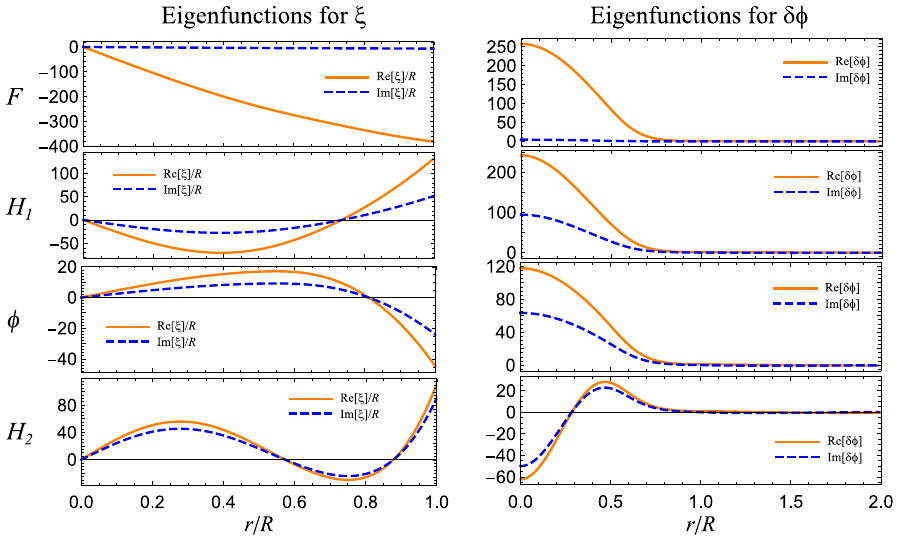} 
    \caption{Real (orange) and imaginary (dashed blue) parts of the eigenfunctions for the Lagrangian displacement $\xi$ and scalar field perturbation $\delta \phi$ for the four lower frequency modes of a scalarized NS with $\tilde{\rho}_c = 11.65 \rho_0$ in the MO100 model.}
    \label{fig:MO_eigenfunctions}
\end{figure*}

Let us now turn to the MO100 model. 
Figure \ref{fig:linearspectrumMO} shows frequency as a function of central density for this model, where one can identify the fundamental scalar-led ($\phi$) mode in addition to the ($F$, $H_1$, and $H_2$) fluid-led modes. As discussed before, the frequency of the $\phi$-mode goes to zero at the boundary of the scalarization region. However, mode identification after the onset of scalarization is less straightforward in this case\footnote{In Ref.~\cite{Mendes2018}, the branch continuously connected to the $\phi$-mode was identified as belonging also to the $\phi$-mode class. However, the present analysis supports the conclusion that the three lower-frequency branches of modes change their nature as a result of avoided crossings.}.
In order to help us identify the nature of the various branches shown in Fig.~\ref{fig:linearspectrumMO}, we represent with a dashed gray line the fundamental frequency of scalar field perturbations computed in an ``inverse Cowling'' approximation, i.e.~by freezing fluid perturbations. Since the coupling between scalar and fluid perturbations is small, specially near the critical central density for spontaneous scalarization, we expect the $\phi$-mode frequency not to differ considerably from that obtained within the inverse Cowling approximation. This expectation, along with the behavior seen in Fig.~\ref{fig:linearspectrumMO},
supports the conclusion that the nature of the various continuous branches changes as central density increases, as a result of avoided crossings between these mode frequencies. In particular, the fact that the size of the gap at the second avoided crossing (between the $\phi$ and $H_1$-mode) is larger than the first (between the $\phi$ and $F$-mode) is expected as a consequence of the weaker coupling at the first avoided crossing. Therefore, slightly after the onset of scalarization, the $\phi$-mode is not anymore the fundamental mode of the scalarized solution, with that role being played by the fluid-led $F$-mode. Additionally, as a result of the weak coupling between fluid and scalar field perturbations, energy is not as efficiently transferred to scalar radiation as in the DEF-6 model, and the damping timescale $\tau_\text{damp} = \Im(\omega)^{-1}$ is typically larger. For the maximum mass solution, $\tau_\text{damp} \approx 0.11$ s for the $\phi$-mode, and even larger for the other modes shown in Fig.~\ref{fig:linearspectrumMO}.

\begin{table*}[bht]
  \centering
 \begin{tabular}{|c | c | c | c | c | c | c | c | c | c | c | } 
 \hline
 Simulation & Model & $\beta$ & $\tilde{\rho}_c/\rho_0$ (initial) & $|\phi_c|$ (initial) & $\Delta \tilde{\rho}_c$ & $M_b (M_\odot)$ (after perturb.)& $\tilde{\rho}_c / \rho_0$ (final) & $|\phi_c|$ (final) \\ [0.5ex] 
 \hline
$S_\text{DEF}$ & DEF & $-6$ & 10.0 & 0 & - & $2.362 \pm 0.003$ & 8.080 & 0.2838   \\
 \hline 
$S_\text{MO}$ & MO & 100 & 11.2 & 0 & - & $2.432 \pm 0.001$ & 11.65 & 0.02734  \\
\hline
$M_\text{DEF}$ & DEF & $-6$ & 17.031 & 0 & 5\%  & $2.244 \pm 0.014$ & 7.918 & 0.2800  \\ 
 \hline
$M_\text{MO}$ & MO & 100 & 14.0 & 0 & 2\% & $2.406 \pm 0.04$ & 10.71 & 0.01052  \\
 \hline
$MS_\text{DEF}$ &  DEF & $-6$ & 10.0 & 0.1949 & 1\% & $2.931 \pm 0.008$ & 8.860 & 0.2839 \\ 
 \hline
$MS_\text{MO}$ & MO & 100 & 12.228 & 0.03904 & 1.6\% & $2.399 \pm 0.003 $ & 10.57 & 0.007718   \\
[1ex] 
 \hline
\end{tabular}
  \caption{Data characterizing simulations: (i) simulation identifier, (ii) model, (iii) $\beta$, (iv) central density and (v) central value of the scalar field for the initial equilibrium solution (before density perturbation), (vi) density perturbation (in migration experiments), (vii) baryon mass after perturbation (error bars take into account the evolution of the baryon mass during the simulation), (viii) central density and (ix) central scalar field of the stable equilibrium solution with baryon mass given by the mean value in column (vii).}
  \label{tab:simulations}
\end{table*}

Figure \ref{fig:MO_eigenfunctions} shows the eigenfunctions for the $F$, $H_1$, $\phi$, and $H_2$ modes of a scalarized solution with central density $\tilde{\rho}_c = 11.65 \rho_0$ (the central density of the scalarized NS formed in simulation $S_\text{MO}$; see below) in the MO100 model. The Lagrangian displacement has no nodes for the $F$ mode, one node for the $\phi$ and $H_1$ modes, and two nodes for the $H_2$ mode. Indeed, the number of nodes in the Lagrangian displacement for the $\phi$-mode changes as central density increases, as it becomes influenced by different fluid modes across avoided crossings. The small damping found in the MO100 model reflects itself in a much slower growth of the eigenfunctions for $\delta \phi$ as $r \to \infty$, when we compare Figs.~\ref{fig:DEF_eigenfunctions} and \ref{fig:MO_eigenfunctions}.

\section{Nonlinear dynamics of scalarized solutions} \label{sec:nonlinear}

\subsection{Description of the simulations}

In this section we describe the main results of this work. We explore the nonlinear dynamics of scalarized solutions formed in three different scenarios, described pictorially in Fig.~\ref{fig:types}. Scenario of type $S$ features the process of spontaneous scalarization: The initial configuration is a trivial, GR-like solution that is unstable under scalar field perturbations. Scenarios of type $M$ and $MS$ feature migrations, to the stable branch, of initial solutions that are unstable under fluid perturbations, i.e., lying beyond the turning point in the $M_b-\tilde{\rho}_c$ diagram of Fig.~\ref{fig:equilibrium}. The initial configuration may be scalarized---corresponding to class $MS$---, or not---corresponding to class $M$. 
Table \ref{tab:simulations} condenses relevant information about the six simulations analyzed in what follows, including details of the initial condition.
As discussed in Sec.~\ref{sec:setup_numerics}, we perturb the initial equilibrium solution in simulations type $M$ and $MS$ to force the evolution towards a stable, lower density configuration and avoid collapse to a black hole. The perturbation strength in each case is indicated in Table \ref{tab:simulations}, as well as the baryon mass after perturbation. This quantity should be conserved during the evolution, but it changes slightly due to numerical errors---see Sec.~\ref{sec:setup_numerics}. In Table~\ref{tab:simulations}, the mean value of the baryon mass is reported, with errors indicating the range covered in the simulation. The final equilibrium configuration, with central density and central scalar field indicated in Table \ref{tab:simulations}, is assumed to be the one with the mean baryon mass inferred from the simulation data; these solutions are also represented in Fig.~\ref{fig:equilibrium}.
Note, however, that different prescriptions to identify the final equilibrium solution are possible, since baryon mass is not strictly conserved during the simulation, and this would slightly alter the mode frequencies inferred from perturbation theory and used to interpret our results.

In what follows, we analyze each simulation separately. We begin with simulation class $S$.

\begin{figure}[thb]
    \includegraphics[width=8.5cm]{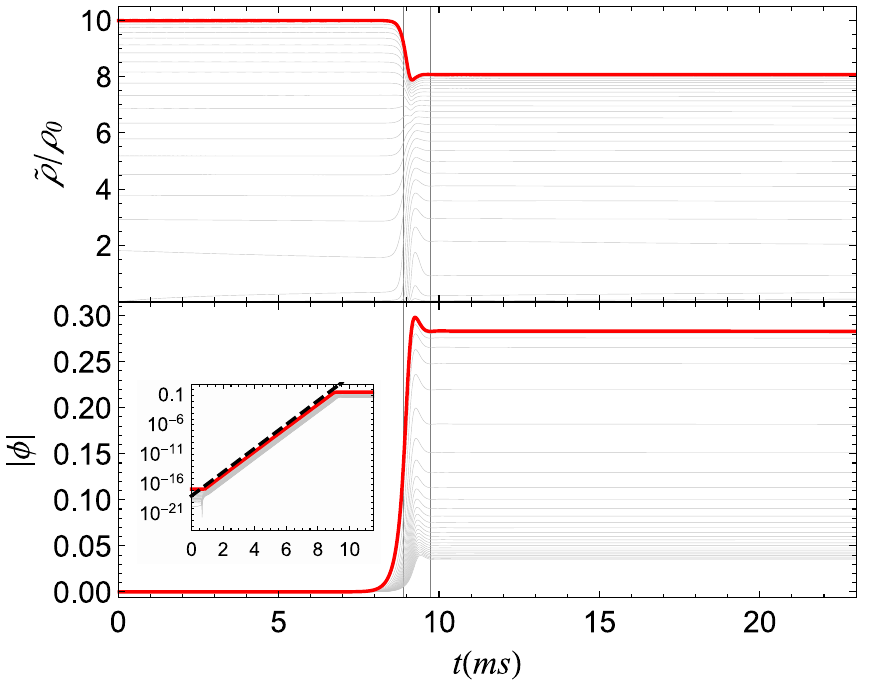} 
    \caption{Time evolution of rest-mass density and scalar field in simulation $S_\text{DEF}$. Each gray curve corresponds to data extracted at a fixed spatial point inside the star, with the red curves corresponding to $r=0$. The inset of the lower panel shows the scalar field evolution in the first milliseconds in a log-scale; the black dashed curve is proportional to $e^{t/\tau}$, where $\tau = 0.217$ ms is the timescale for the instability of the initial solution under scalar field perturbations, as predicted by linear perturbation theory (see Fig.~\ref{fig:instability}). Vertical lines highlight the values 8.90 ms and 9.77 ms.}
    \label{fig:S_DEF_time_evol}
\end{figure}

\begin{figure}[th]
    \centering
    \includegraphics[width=8cm]{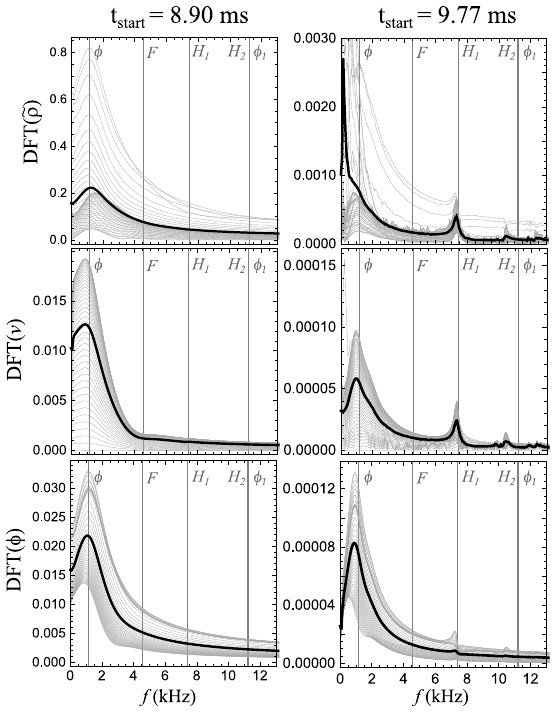}
    \caption{Discrete Fourier Transform of the rest-mass density (upper panels), radial velocity (middle panels) and scalar field (lower panels), starting at 8.90 ms (left column) and 9.77 ms (right column), for simulation $S_\text{DEF}$. Gray curves are DFTs corresponding to data extracted at different fixed radial positions, and the black curve represents the spatial average. Gray vertical lines indicate the mode frequencies obtained from perturbation theory: $\omega_{\phi}/(2\pi) = (1.16 + 0.99 i) \textrm{ kHz}$,  $\omega_{F}/(2\pi) = (4.54 + 1.63 i) \textrm{ kHz}$, $\omega_{H_1}/(2\pi) = (7.42 + 0.15 i) \textrm{ kHz}$, $\omega_{H_2}/(2\pi) = (11.2 + 0.13 i) \textrm{ kHz}$, and $\omega_{\phi_1}/(2\pi) = (11.2 + 2.72 i) \textrm{ kHz}$.}
    \label{fig:S_DEF_dft}
\end{figure}

\begin{figure}[htb]
    \centering
    \includegraphics[width=7.5cm]{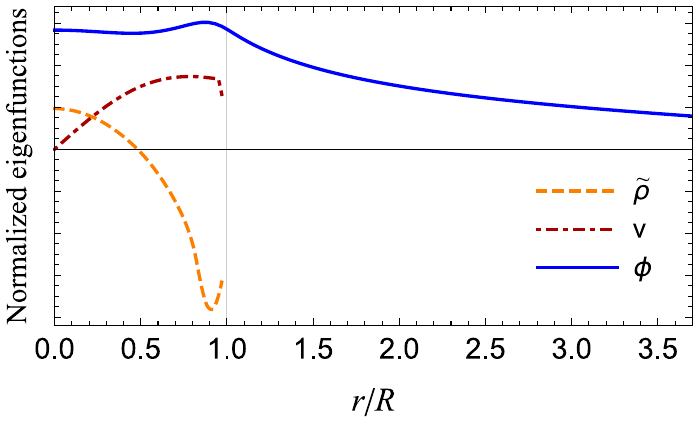}
    \caption{Normalized eigenfunctions for the rest-mass density, radial velocity and scalar field, reconstructed from simulation $S_\text{DEF}$, for $t> 8.90$ ms, at a fixed frequency $f = 1.16$ kHz. The reconstructed eigenfunctions are consistent with expectations from linear perturbation theory (cf.~Fig.~\ref{fig:DEF_eigenfunctions}).}
    \label{fig:S_DEF_mode_reconstruction}
\end{figure}

\subsection{Type $S$}

\subsubsection{$S_\text{DEF}$: Spontaneous scalarization in the DEF-6 model}

Simulation $S_\text{DEF}$ features the process of spontaneous scalarization in the DEF model, which has been studied extensively in the literature since the initial numerical explorations \cite{Novak1998}. Figure \ref{fig:S_DEF_time_evol} shows the time evolution of the rest-mass density and scalar field extracted at different radial positions. The initial configuration lies in the GR branch of Fig.~\ref{fig:equilibrium}, and has an almost trivial scalar field profile, $\phi \approx \phi_0 = 0$. In practice, the initial value for the scalar field is determined by round-off errors; physically, small deviations from $\phi_0 = 0$ are to be expected, even if of quantum-mechanical origin \cite{Lima2010,Mendes2014}. This small scalar seed undergoes a phase of exponential amplification, as visible in the inset of the lower panel of Fig.~\ref{fig:S_DEF_time_evol}. The timescale for the exponential growth predicted by linear perturbation theory ($\tau = 0.217$ ms; see Fig.~\ref{fig:instability}) agrees well with the numerical data: In this initial phase, the scalar field behaves as a linear perturbation evolving on a fixed background. Around $t=9$ ms, the scalar field has grown sufficiently large and we see a quick convergence of the scalar and density profiles to the final scalarized equilibrium solution.

\begin{figure}[b]
    \includegraphics[width=8.5cm]{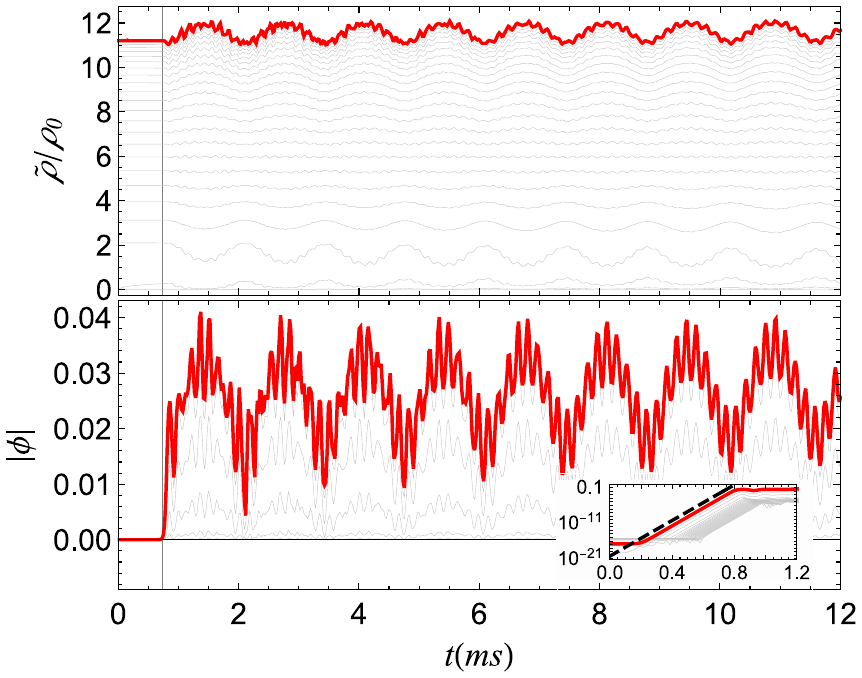} 
    \caption{Same as Fig.~\ref{fig:S_DEF_time_evol}, but for simulation $S_\text{MO}$. In the inset of the lower panel, the black dashed curve is proportional to $e^{t/\tau}$, where $\tau = 0.0174$ ms is the timescale for the (linear) instability of the initial solution under scalar field perturbations (see Fig.~\ref{fig:instability}). The vertical line highlights the value $0.73$ ms.}
    \label{fig:S_MO_time_evol}
\end{figure}

Figure~\ref{fig:S_DEF_dft} shows the DFT of the rest-mass density, radial velocity and scalar field, starting at two different times ($t_\text{start}$) marked as vertical lines in Fig.~\ref{fig:S_DEF_time_evol}. Frequencies for the $\phi$, $F$, $H_1$, $H_2$, and $\phi_1$ modes obtained from the linear analysis (cf.~Sec.~\ref{sec:linear}) are also represented in the plot. Figure \ref{fig:S_DEF_mode_reconstruction} shows the wave function reconstructed from the numerical data at peak frequency, which agrees with the expectation for the $\phi$-mode according to the upper-right panel of Fig.~\ref{fig:DEF_eigenfunctions}. Thus, we see that the evolution towards the scalarized state is driven by the $\phi$-mode. This mode decays quickly, radiating away most of the energy contained in the system. At later times, higher order, less damped modes become more important, specially the $H_1$-mode. 
\begin{figure}[th]
    \subfloat{{\includegraphics[width=7.5cm]{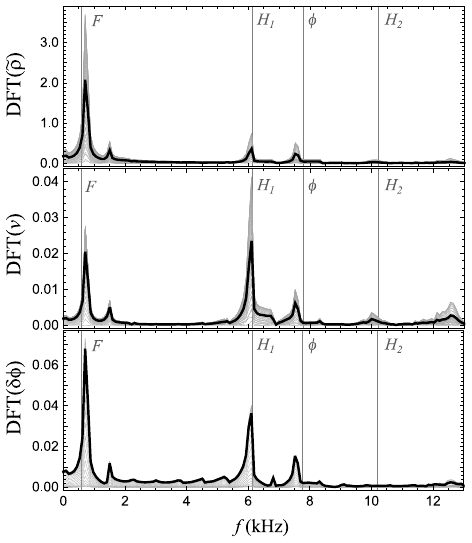}}}
    \caption{Discrete Fourier Transform of rest-mass density, radial velocity and scalar field, for simulation $S_\text{MO}$, starting at 0.73 ms. Gray curves indicate the DFT of the corresponding variables at fixed spatial points, while the black curve represents an average over the star. Gray vertical lines correspond to frequencies obtained from linear perturbation theory: $\omega_F/(2\pi) = (0.574 + 6.7 \times 10^{-6} i) \textrm{ kHz}$,  $\omega_{H_1}/(2\pi) = (6.12 + 2.2 \times 10^{-4} i) \textrm{ kHz}$, $\omega_{\phi}/(2\pi) = (7.77 + 1.3 \times 10^{-3} i) \textrm{ kHz}$, $\omega_{H_2}/(2\pi) = (10.2 + 6.7 \times 10^{-4} i) \textrm{ kHz}$.}
    \label{fig:S_MO_DFT}
\end{figure}

\begin{figure}[th]
    \subfloat{{\includegraphics[width=7.5cm]{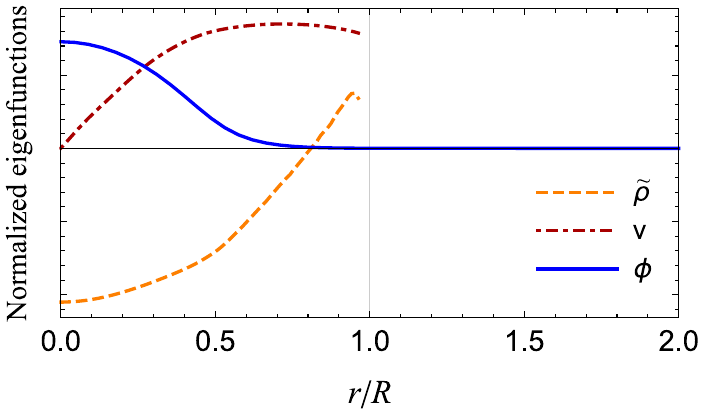}}}
    \caption{Normalized eigenfunctions for rest-mass density, radial velocity and scalar field, reconstructed from the numerical data for simulation $S_\text{MO}$ at 1.5 kHz. They support the interpretation that the peak at 1.5 kHz is due to a non-linear self-coupling of the $F$-mode.
}
    \label{fig:S_MO_reconstruction}
\end{figure}

\subsubsection{$S_\text{MO}$: Spontaneous scalarization in the MO100 model}

Simulation $S_\text{MO}$ again features the process of spontaneous scalarization, but now in the MO model \cite{Mendes2016}. The time evolution of the rest-mass density and scalar field is shown in Fig.~\ref{fig:S_MO_time_evol}. Two main differences can be noticed from the previous case ($S_\text{DEF}$): First, the initial phase of exponential growth is much shorter, as a result of the typically smaller instability timescales in the MO100 model (see Fig.~\ref{fig:instability}). Second, large amplitude oscillations around the final equilibrium solution can be seen during the entire simulation time span. This reflects the fact that the evolution is now driven by long lived modes. As a result of these persistent large amplitude oscillations, nonlinear features are more likely to appear in the spectrum.

Figure \ref{fig:S_MO_DFT} shows the DFT of the rest-mass density, radial velocity and scalar field, starting at $t_\text{start} = 0.73$ ms (indicated as a vertical line in Fig.~\ref{fig:S_MO_time_evol}). The spectrum shows clear peaks corresponding to the $F$, $H_1$, and $\phi$ modes, which are responsible for the low and high frequency components present in Fig.~\ref{fig:S_MO_time_evol}. The eigenfunctions reconstructed from the numerical data at these frequencies are consistent with those shown in Fig.~\ref{fig:MO_eigenfunctions}, supporting the mode identification. 

\begin{figure}[b]
    \centering
    {\includegraphics[width=8.5cm]{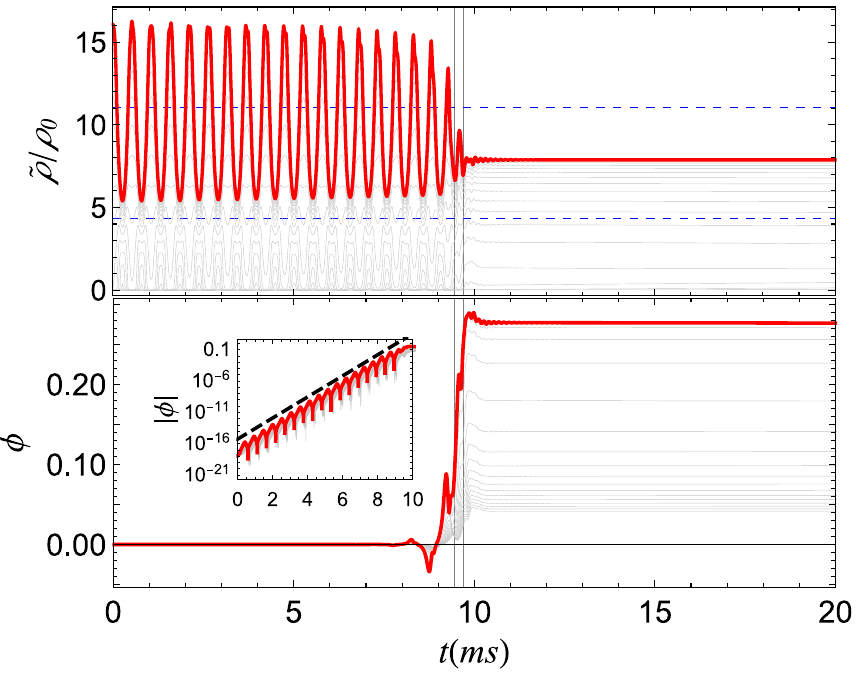} }
    \caption{Time evolution of rest-mass density and scalar field in simulation $M_\text{DEF}$. Each gray curve corresponds to data extracted at a fixed spatial point inside the star, with the red curves corresponding to $r=0$. Vertical lines denote $t = 9.45$ ms and $t = 9.70$ ms. Two dashed blue horizontal lines in the top panel indicate the range of central densities where scalarized solutions exist. The black dashed line in the inset of the bottom panel is proportional to $e^{t/\tau}$, with $\tau = 0.259$ ms.}
    \label{fig:M_DEF_time_evol}
\end{figure}

\begin{figure*}[tbh]
    \centering
    {\includegraphics[width=17cm]{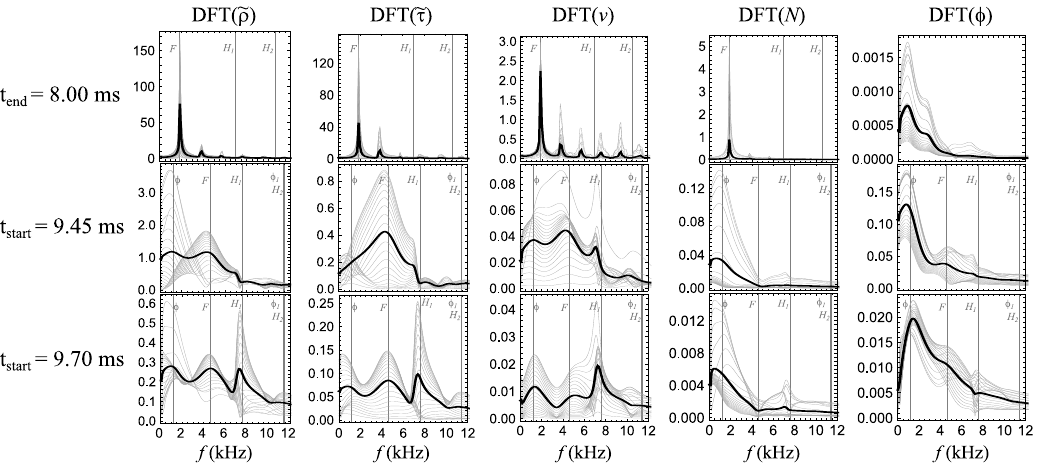} }
    \caption{Discrete Fourier Transform of rest-mass density $\tilde{\rho}$, energy density $\tilde{\tau}$, radial velocity $v$, lapse $N$, and scalar field $\phi$, for simulation $M_\text{DEF}$. Gray curves indicate the DFT of the corresponding variables at fixed spatial points, while the black curve represents a spatial average. 
    \textit{Top row}: DFT taken from $t_\text{start} = 0.0$ ms to $t_\text{end} = 8.0$ ms. Vertical lines indicate the $F$, $H_1$, and $H_2$ mode frequencies of a star with $\tilde{\rho}_c = 10.23 \rho_0$ in GR. 
    \textit{Middle and bottom rows}: DFT taken from $t_\text{start} = 9.45$ ms (middle) and $t_\text{start} = 9.70$ ms (bottom) up to $t_\text{end} = 20.0$ ms. Vertical lines correspond to mode frequencies of the final scalarized solution, obtained from perturbation theory. These frequencies are: $\omega_{\phi}/(2\pi) = (1.20 + 1.07 i) \textrm{ kHz}$,  $\omega_{F}/(2\pi) = (4.62 + 1.48 i) \textrm{ kHz}$, $\omega_{H_1}/(2\pi) = (7.59 + 0.195 i) \textrm{ kHz}$, $\omega_{H_2}/(2\pi) = (11.43 + 0.151 i) \textrm{ kHz}$, and $\omega_{\phi_1}/(2\pi) = (11.38 + 2.83 i) \textrm{ kHz}$, and the corresponding eigenfunctions are qualitatively similar to those presented in Fig.~\ref{fig:DEF_eigenfunctions}.}
    \label{fig:M_DEF_dft}
\end{figure*}

Interestingly, the spectrum reveals additional features that are not explained by linear perturbation theory alone. In particular, the DFTs for all fluid and space-time variables display a clear peak at approximately 1.5 kHz. This corresponds to roughly twice the frequency of the main peak in the density and scalar field spectra (at $\sim0.71$ kHz), ascribed to the fluid-led $F$-mode. The eigenfunctions extracted at this frequency are shown in Fig.~\ref{fig:S_MO_reconstruction}. The fact that the wave-function for the rest-mass density has a single node and the one for the radial velocity field has no nodes is consistent with the interpretation that the peak at 1.5 kHz comes from a non-linear self-coupling of the $F$-mode. As we will see, this feature will be present in all  simulations in the MO model.

\subsection{Type $M$}

We proceed to analyze simulations of type $M$, which represent the formation of a stable, scalarized solution from an initial configuration that is unstable under gravitational collapse---a migration from the unstable branch of equilibrium solutions (beyond the turning point) to the stable branch. The initial solution has zero scalar charge; the case where the scalar charge is nonzero initially corresponds to the simulation class $MS$ and will be analyzed in the next subsection.

\subsubsection{$M_\text{DEF}$: Migration in the DEF-6 model}

Figure \ref{fig:M_DEF_time_evol} shows the time evolution of the rest-mass density and scalar field at equally spaced positions, for simulation $M_\text{DEF}$. Two phases can be clearly identified. 

In the first phase, the scalar field has a small amplitude and the stellar fluid behaves as in GR, undergoing large amplitude oscillations. The mean central density in the first phase is approximately $10.23 \rho_0$---note that this is somewhat higher than the central density of a GR equilibrium solution with the baryon mass of simulation $M_\text{DEF}$ ($2.244 M_\odot$), which would have $\tilde{\rho}_c \approx 8.92 \rho_0$.  Along with fluid oscillations, the small scalar seeds present in the initial data as numerical errors undergo a phase of exponential growth modulated by oscillations. The timescale for the exponential growth is well approximated by $\tau = 0.259$ ms, as seen in the inset of the bottom panel of Fig.~\ref{fig:M_DEF_time_evol}; this is precisely the instability timescale for scalar field perturbations around a GR equilibrium solution with $\tilde{\rho}_c = 10.23 \rho_0$ (cf.~Fig.~\ref{fig:instability}). 

The oscillation spectrum for this first phase can be seen in the top row of Fig.~\ref{fig:M_DEF_dft}, which shows the DFT of various fluid and spacetime quantities -- rest-mass density, energy density, radial velocity, lapse, and scalar field --, taken from $t_\text{start} = 0.0$ ms up to $t_\text{end} = 8.0$ ms. The star oscillates predominantly at $\sim 1.87$ kHz, which is very close to the frequency of the fundamental radial mode of an equilibrium solution with $\tilde{\rho}_c = 10.23 \rho_0$ --- the mean central density in this phase. The $H_1$ and $H_2$ modes are not significantly excited. However, the spectrum reveals additional peaks at frequencies corresponding to multiples of the $F$-mode frequency, likely due to the non-linear self-coupling of the $F$-mode.

While the GR-like star oscillates in this manner, driven by its $F$-mode, the scalar field undergoes an exponential growth, modulated by oscillations. Since the scalar field amplitude is small, one would expect scalar field perturbations to approximately decouple from fluid perturbations (cf.~Sec.~\ref{sec:radialeqs}). Indeed, the scalar field spectrum shown in the last panel of the top row of Fig.~\ref{fig:M_DEF_dft} is radically different from the other spectra of the same row, with a broad peak at $\sim 0.875$ kHz. During this initial phase, scalar perturbations respond to a nontrivial, rapidly evolving background, and their spectrum is not characteristic of the ``mean'' equilibrium solution.

The second phase in simulation $M_\text{DEF}$ is characterized by a quick convergence to the scalarized equilibrium configuration, due to the excitation of the strongly damped modes of that final solution.
The middle and bottom rows of Fig.~\ref{fig:M_DEF_dft} show the DFT for this phase, starting at $t_\text{start} = 9.45$ ms (middle row) and 9.70 ms (bottom row). The spectrum is richer than that of simulation $S_\text{DEF}$ showing spontaneous scalarization in the same model (see Fig.~\ref{fig:S_DEF_dft}). There, the evolution was driven by the fundamental scalar-led mode, while in this migration experiment the three lowest frequency modes are substantially excited; they correspond to the fundamental scalar-led mode and the two lowest-frequency fluid-led modes. The associated peaks are broad, due to the large imaginary part of the mode frequencies, reflecting their strong damping in time. No peaks corresponding to nonlinear mode couplings are clearly identifiable.

\subsubsection{$M_{MO}$: Migration in the MO model}

\begin{figure}[htb]
    \includegraphics[width=8.5cm]{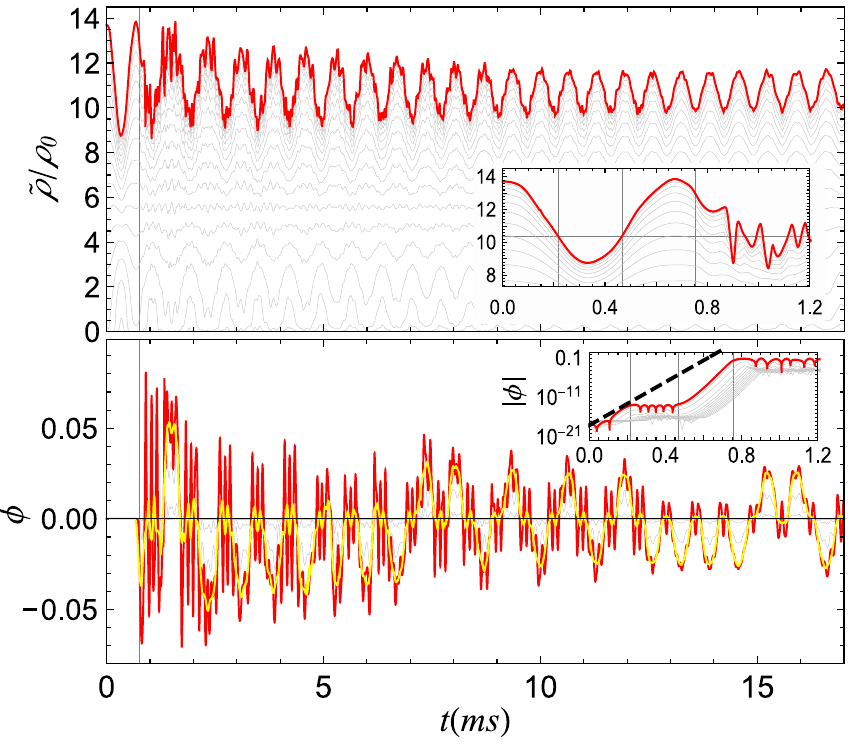}
    \caption{Time evolution of the rest-mass density and scalar field in simulation $M_\text{MO}$. Each gray curve corresponds to data extracted at different fixed spatial points, and the red curves corresponds to $r=0$. The insets highlight the first 1.2 ms of the simulation, with vertical lines at $t = 0.22$ ms, $t = 0.47$ ms, and $t = 0.76$ ms, which mark qualitative changes in the behavior of the scalar field. A horizontal line at the inset of the upper panel highlights the value $\tilde{\rho} = 10.4 \rho_0$: a GR equilibrium solution with central density above this value is unstable under scalar field perturbations in the MO100 model. The black dashed line in the inset of the bottom panel is proportional to $e^{t/\tau}$, where $\tau = 0.0158$ ms is the timescale for the linear instability of the initial solution under scalar field perturbations. In the bottom panel, the yellow curve represents a moving average of the central value of the scalar field.}
    \label{fig:M_MO_time_evol}
\end{figure}

Next, we proceed to analyze simulation $M_\text{MO}$, which displays a rich and complex behavior. Figure \ref{fig:M_MO_time_evol} shows the time evolution of the rest-mass density and scalar field at various spatial points. Again, two phases can be identified.

In the first phase, that lasts for the first $\sim 0.8$ ms, the scalar field is small and behaves as a linear perturbation in the background of an oscillating general-relativistic star. The mean central density in this first phase is approximately $11.3 \rho_0$. 

Contrary to the previous simulations, the first phase is not characterized by a sustained exponential growth, but instead can be divided into three pieces, as shown in the insets in Fig.~\ref{fig:M_MO_time_evol}. In the first ($t \lesssim 0.22$ ms) and third ($0.47 \textrm{ ms} \lesssim t \lesssim 0.76 \textrm{ ms}$) pieces, the scalar field grows exponentially. The timescale for the exponential growth is relatively well approximated by the instability timescale of the initial solution, which is 0.0158 ms for a star with $\tilde{\rho}_c = 13.72 \rho_0$---that is the initial central density after perturbation (see Table~\ref{tab:simulations}).

However, the exponential growth is halted for $0.22 \textrm{ms} \lesssim t \lesssim 0.47 \textrm{ms}$. In this time interval, the background solution has a central density below $10.4 \rho_0$, which is the critical value for the appearance of unstable modes in the MO100 model. Thus, for a fraction of a millisecond, while the star rebounds to become denser again, the instability is quenched and the scalar field performs a few oscillations, with a frequency of $\sim$ 12 kHz. This is consistent with what we would expect from perturbation theory: An equilibrium configuration with $\tilde{\rho}_c = 9.33 \rho_0$ (which is the mean central density for $0.22 \textrm{ms} \lesssim t \lesssim 0.47 \textrm{ms}$) has a pure scalar mode with frequency $\omega_\phi/(2\pi) = (11.1 + 0.000970 i)$ kHz. 

Note that in simulation $M_\text{DEF}$, the star also makes excursions outside of the instability band, but there is no pause in the exponential growth of the scalar field (see Fig.~\ref{fig:M_DEF_time_evol}). The reason for these distinct behaviors has to do with the interplay between two relevant timescales: one governing the growth of scalar perturbations (associated with the unstable $\phi$-mode frequency), other governing the stellar oscillation (associated with its $F$-mode frequency). In simulation $M_\text{MO}$, the background oscillates slowly in comparison with the timescale for the scalar growth. Indeed, the fundamental mode for an equilibrium solution with $\tilde{\rho}_c = 11.3 \rho_0$ (the mean central density in the first phase) has a frequency of 1.41 kHz, and thus a typical timescale of 0.71 ms, which is a much longer timescale than the one associated with the exponential growth of scalar field perturbations ($\sim$ 0.0158 ms). On the other hand, these timescales are comparable in simulation $M_\text{DEF}$.

\begin{figure}[thb]
    \centering
    \includegraphics[width=7.8cm]{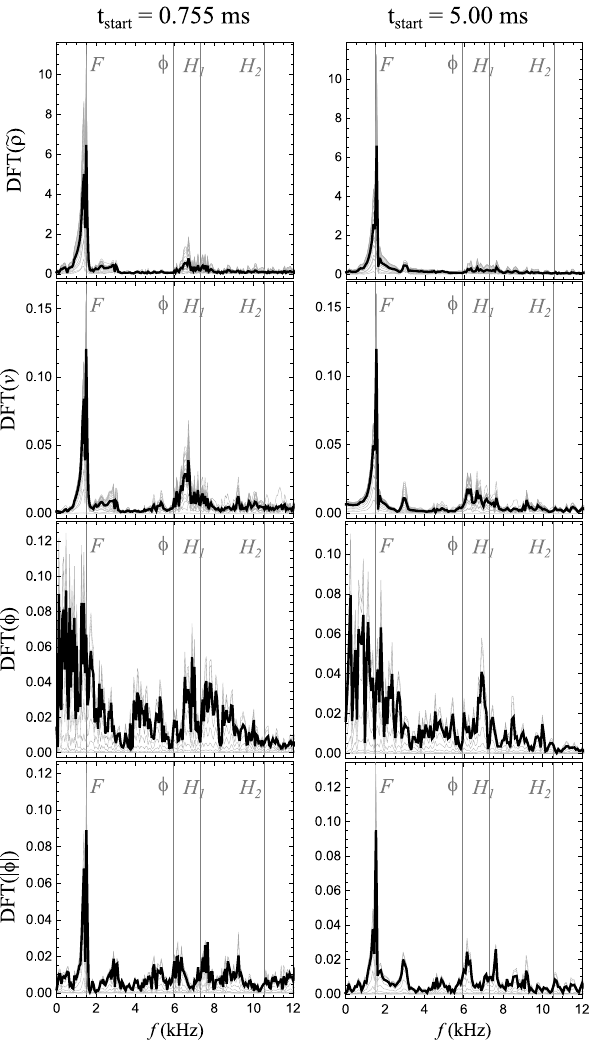}
    \caption{Discrete Fourier Transform of the rest-mass density, radial velocity, scalar field, and absolute value of the scalar field, for simulation $M_\text{MO}$, starting at $t_\text{start} = 0.755$ ms (left column) and $t_\text{start} = 5.00$ ms (right column). Gray curves represent the DFT of the corresponding variables taken at different fixed points inside the star, while the black curve corresponds to a spatial average. Gray vertical lines show linear mode frequencies of the stable equilibrium solution with the total baryon mass present in the simulation. These are: $\omega_F/(2\pi) = (1.49 + 2.53 \times 10^{-6} i) \textrm{ kHz}$, $\omega_\phi/(2\pi) = (5.91 + 2.91 \times 10^{-4} i) \textrm{ kHz}$, $\omega_{H_1}/(2\pi) = (7.28 + 2.29 \times 10^{-4} i) \textrm{ kHz} $, and $\omega_{H_2}/(2\pi) =  (10.55 + 1.32 \times 10^{-4} i) \textrm{ kHz}$.}
    \label{fig:M_MO_dft}
\end{figure}

After the first phase described above, the scalar field starts to influence the overall evolution. From Fig.~\ref{fig:M_MO_time_evol}, the rest-mass density is seen to acquire higher frequency modulations that are present in the scalar field spectrum, and its oscillations are (slowly) damped due to scalar radiation. Since the $F$-mode frequency of the final scalarized solution does not differ appreciably from that of GR (cf.~Fig.~\ref{fig:linearspectrumMO}), there is no considerable change in the main, low frequency component. 

Before analyzing the spectrum in more detail, it is interesting to notice that the scalar field evolution, shown in the bottom panel of Fig.~\ref{fig:M_MO_time_evol}, is much noisier than in simulation $S_\text{MO}$, which features the process of spontaneous scalarization in the same model. This has to do with the fact that the scalar field executes large-amplitude oscillations, in a range that encompasses the two possible equilibrium solutions it may settle to (with a central value $\phi_c \approx \pm 0.01$). The bottom panel of Fig. \ref{fig:M_MO_time_evol} includes a yellow curve representing the moving average of the central value of the scalar field: Analyzing this curve we can see more clearly that the scalar field alternates between oscillating around the negative-valued and the positive-valued solutions. This generates a more complex pattern that is reminiscent of the evolution of chaotic systems with Lorenz attractors~\cite{Devaney2003}.

Figure \ref{fig:M_MO_dft} shows the DFT for the second phase of the evolution, starting at $t_\text{start} = 0.755$ ms (left column) and $t_\text{start} = 5.00$ ms (right column). The spectrum becomes cleaner as time passes. As a result of the large excursions of the scalar field, oscillating around the two possible equilibrium solutions, the spectrum for the scalar field (third row of Fig.~\ref{fig:M_MO_dft}) is extremely noisy, with no clear peaks discernible, but the situation changes when we look at its absolute value (fourth row of Fig.~\ref{fig:M_MO_dft}). For all variables, the fluid-led $F$-mode is the most excited. The $\phi$ and $H_1$ modes are also discernible. Note, from Fig.~\ref{fig:linearspectrumMO}, that the central density of the final scalarized solution, $\tilde{\rho}_c = 10.71 \rho_0$, is close to the avoided crossing between the $\phi$ and $H_1$ modes, and mode identification is less obvious in this region. We can also distinguish a peak around 3 kHz, which is roughly twice the $F$-mode frequency, and that can be ascribed to the nonlinear self-coupling of this mode. Analysis of the wavefunction reconstructed at this frequency corroborates this interpretation.

\subsection{Type $MS$}

Finally, we discuss two migration experiments, where the initial data is a (perturbed) unstable equilibrium solution, but which already presents a nontrivial scalar charge. The main difference we will see regards the first phase of the evolution, which in previous simulations was characterized by the exponential growth of scalar perturbations in a background that evolved unaffected by them. This phase will not be present in simulations of type $MS$, since fluid and scalar field are strongly coupled from the very beginning.

\subsubsection{$MS_\text{DEF}$: Migration from scalarized in the DEF model}

\begin{figure}[t]
    \centering
    \includegraphics[width=8cm]{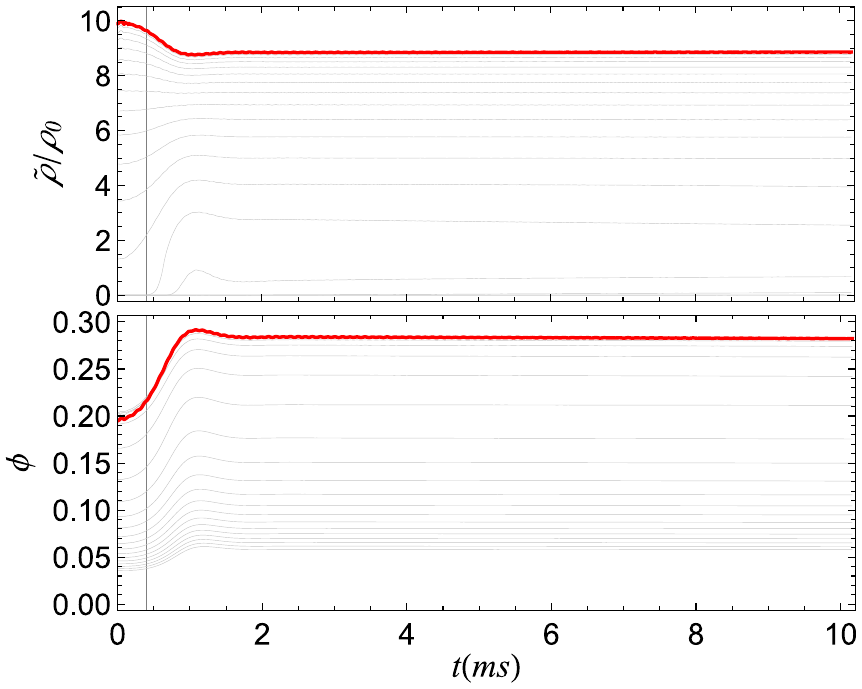}
    \caption{Time evolution of the rest-mass density and scalar field in simulation $MS_\text{DEF}$. Each gray curve corresponds to data taken at different fixed spatial points, and the red curves corresponds to $r=0$. A vertical line is shown at $t = 0.40$ ms.}
    \label{fig:MS_DEF_time_evol}
\end{figure}

This simulation is similar in spirit to simulation $M_\text{DEF}$, but starting from a scalarized initial configuration. Figure \ref{fig:MS_DEF_time_evol} shows the time evolution of the rest-mass density and scalar field, where we can see a smooth transition to the new, stable scalarized solution. 

\begin{figure}[th]
    \centering
    \includegraphics[width=8cm]{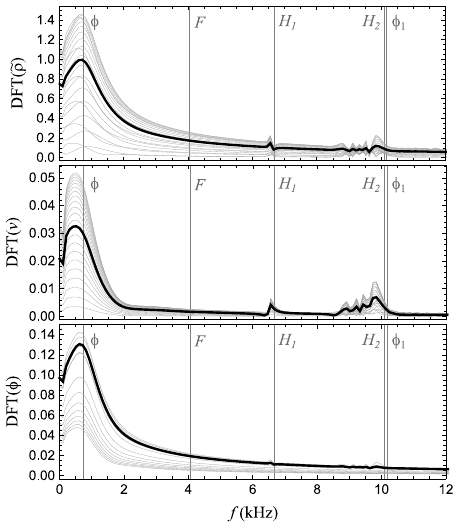}
    \caption{Discrete Fourier Transform of the rest-mass density, radial velocity, and scalar field, for simulation $MS_\text{DEF}$, starting at $t_\text{start} = 0.40$ ms. Gray curves represent the DFT of the corresponding quantities evaluated at fixed spatial points, while black curves correspond to averages over the star. Vertical lines are shown at mode frequencies of the final equilibrium solution, obtained from perturbation theory. These frequencies are $\omega_{\phi}/(2\pi) = (0.745 + 0.529 i) \textrm{ kHz}$, $\omega_F/(2\pi) = (4.06 + 2.02 i) \textrm{ kHz}$, $\omega_{H_1}/(2\pi) = (6.68 + 0.00340 i) \textrm{ kHz}$, $\omega_{H_2}/(2\pi) = (10.09 + 0.0155 i) \textrm{ kHz}$, and $\omega_{\phi_1}/(2\pi) = (10.17 + 2.34 i) \textrm{ kHz}$, and the associated eigenfunctions are qualitatively similar to those presented in Fig.~\ref{fig:DEF_eigenfunctions}.}
    \label{fig:MS_DEF_dft}
\end{figure}

The spectra for the rest-mass density, radial velocity and scalar field are shown in Fig.~\ref{fig:MS_DEF_dft}. As in simulation $S_\text{DEF}$, but in contrast with $M_\text{DEF}$ (where the evolution is governed by the interplay between the first three modes), the evolution is dictated by the $\phi$-mode, with the $H_1$ and $H_2$ modes becoming relevant only at later times, due to their weaker damping.

\subsubsection{$MS_\text{MO}$: Migration from scalarized in the MO model}

\begin{figure}[th]
    \centering
    \includegraphics[width=8cm]{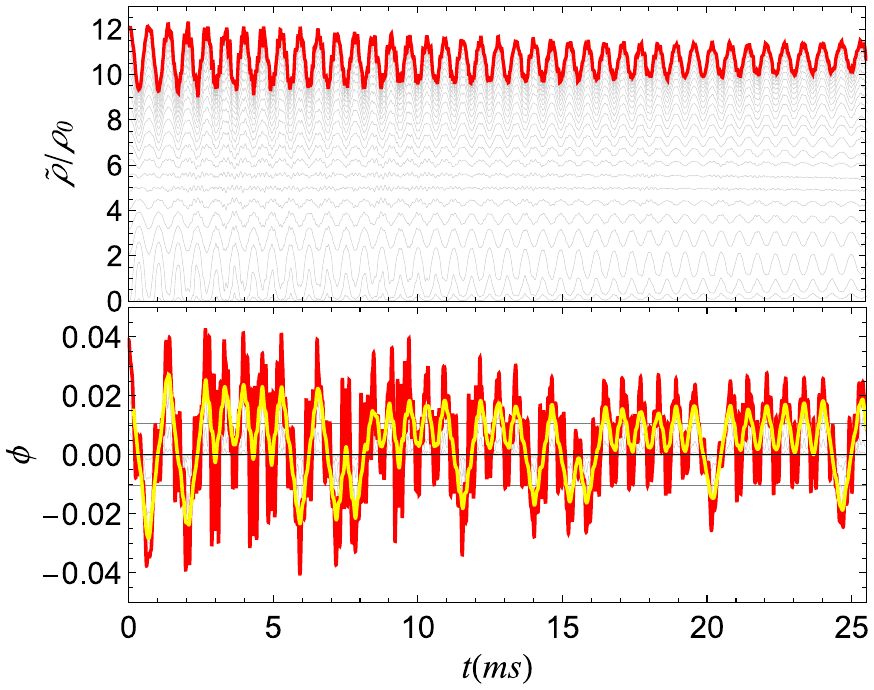}
    \caption{Time evolution of the rest-mass density and scalar field in simulation $MS_\text{MO}$. Each gray curve corresponds to a different fixed spatial point, and the red curves correspond to $r=0$. The yellow curve in the bottom panel shows the moving average of the central value of the scalar field, $\phi_c$.}
    \label{fig:MS_MO_time_evol}
\end{figure}

\begin{figure}[th]
    \centering
    \subfloat{{\includegraphics[width=7.5cm]{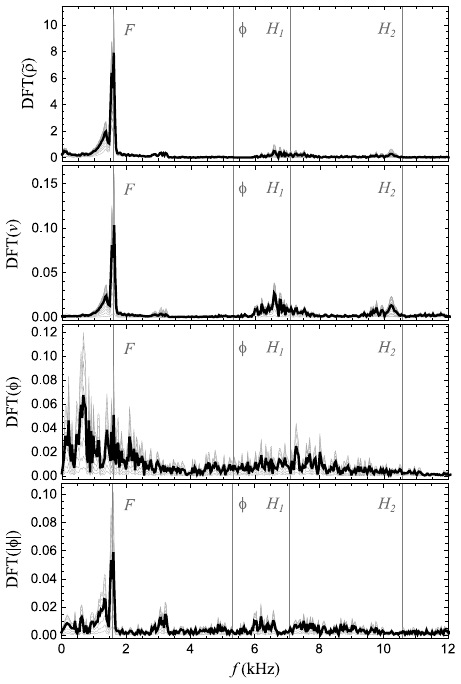}}}
    \caption{Discrete Fourier Transform of the rest-mass density, radial velocity, scalar field, and absolute value of the scalar field for simulation $MS_\text{MO}$. Gray curves represent the DFT of the corresponding quantities evaluated at fixed spatial points, while the black curve corresponds to an average over the star. Vertical lines are shown at the linear mode frequencies of the final equilibrium solution. These are: $\omega_F/(2\pi) =  (1.59 + 2.36 \times 10^{-6} i) \textrm{ kHz}$, $\omega_{\phi} =  (5.31 + 2.77 \times 10^{-4} i) \textrm{ kHz}$, $\omega_{H_1}/(2\pi) = (7.08 + 6.68 \times 10^{-5} i) \textrm{ kHz}$, and $\omega_{H_2}/(2\pi) = (10.58 + 6.79 \times 10^{-5}) \textrm{ kHz}$, and the associated eigenfunctions are qualitatively similar to those presented in Fig.~\ref{fig:MO_eigenfunctions}.}
    \label{fig:MS_MO_dft}
\end{figure}

This simulation is similar to simulation $M_\text{MO}$, but now starting from a scalarized initial configuration. Figure \ref{fig:MS_MO_time_evol} shows the time evolution for the rest-mass density and scalar field. As in simulation $M_\text{MO}$, the range covered by scalar field oscillations encompasses the two possible equilibrium solutions (with positive and negative scalar field profiles). From the moving average shown in the bottom panel of Fig.~\ref{fig:MS_MO_time_evol}, we can see more clearly that the scalar field seems to alternate oscillations around the positive- and negative-valued equilibrium solutions.

The spectra for the rest-mass density, radial velocity, and scalar field are shown in Fig.~\ref{fig:MS_MO_dft}. 
Again, the fluid-led $F$ mode is predominantly excited, with smaller peaks around higher linear mode frequencies and at roughly twice the frequency of the $F$-mode, again showcasing the nonlinear self-coupling of this mode. Due to the large amplitude oscillations of the scalar field, the spectrum is much noisier than in simulation $S_\text{MO}$.

\section{Discussion: Implications for binary neutron star mergers} \label{sec:implications}

The characteristic oscillation modes of NSs carry information about their structure as well as about the underlying theory of gravity. In order to access the potential of GW observations of these modes as a probe of alternative theories of gravity, it is essential to understand the linear and nonlinear dynamics of oscillating NSs in relevant models beyond GR.
In this work we presented results of six $1+1$ nonlinear simulations of (radially) oscillating NSs in two scalar-tensor models, comparing with expectations from perturbation theory. In this section, we discuss the implications of our results for more complex, realistic scenarios.

Naturally, one of the most relevant situations where NS pulsations are significantly excited is the post-merger phase of a binary NS merger, which should be within the observational reach of upgraded or next-generation GW detectors \cite{Bose2018,Yang2018,Martynov2019,Hall2019,Torres-Rivas2019,Ackley2020}. A NS formed as a result of a binary NS merger undergoes strong radial and nonradial oscillations. Although radial oscillations in GR do not couple directly to gravitational radiation, they can influence the GW spectrum. Notably, frequencies corresponding to the quasilinear coupling between the fundamental $l=0$ and $l=2$ modes are seen as secondary, observable peaks in the spectrum coming from $3+1$ numerical relativity simulations.

Contrary to GR, in STTs scalar radiation propagates even in spherical symmetry. However, direct detection of these monopolar scalar waves is hindered by the fact that a detector's response to them is suppressed by factors constrained to be small by solar system tests~\cite{Damour1992,Damour1998}. Even so, one could ask whether, as in GR, radial oscillations may leave imprints in the ringdown GW signal from a binary NS merger. The scalar field could play different roles in such a situation.

First, it is relevant to access whether fluid and scalar field perturbations are coupled since the merger, or whether this coupling develops during the evolution---these scenarios are reminiscent of our simulation classes $MS$ and $M$, respectively. In the classes of STTs studied in this work, coupling between fluid and scalar perturbations is mediated by the background scalar field, which is nontrivial only around scalarized solutions. 
If we consider the merger of two NSs that are not scalarized, forming a scalarized NS, then there will be an initial phase where the scalar field grows exponentially and the fluid evolves independently of it, as in simulations $M_\text{DEF}$ and $M_\text{MO}$ (see Figs.~\ref{fig:M_DEF_time_evol} and \ref{fig:M_MO_time_evol}). The time spent in this initial phase is determined by the instability timescale of the final configuration, which vary among different models (see Fig.~\ref{fig:instability}), as well as by the ambient value of the scalar field, $\phi_0$. The latter is mostly unknown, with quantum fluctuations providing a lower bound \cite{Lima2010,Mendes2014} and solar system constraints providing an upper bound \cite{Will1993}, which may nonetheless be broken in dynamical scenarios \cite{Barausse2013,Palenzuela2014,Sampson2014,Taniguchi2015}. If this initial phase lasts for a time much longer than the NS dynamical timescale, then the evolution should proceed as in GR, and the scalar field effects would appear lately, and may not be observable. 

Second, the coupling strength between fluid and scalar field perturbations needs to be taken into account. The two models analyzed in this work provide radically different examples in this respect. In the DEF-6 model, this coupling is strong; as a result, the kinetic energy present in fluid oscillations is quickly transferred to and radiated away by scalar radiation. Thus we see a quick convergence to the final scalarized solution in all simulations in this model (see~Figs.~\ref{fig:S_DEF_time_evol}, \ref{fig:M_DEF_time_evol}, and \ref{fig:MS_DEF_time_evol}). 

In the DEF-6 model, the GR radial spectrum is substantially modified (see Fig.~\ref{fig:linearspectrumDEF}), with the fundamental scalar-led mode playing a major role in the fluid dynamics. However, due to the strong damping of all radial modes, we do not expect them to have enough time to interact nonlinearly with nonradial modes excited in the post-merger phase of a BNS (which themselves do not seem to couple so strongly to scalar radiation \cite{Kruger:2021yay}).
Therefore, one can speculate that a scalarized solution formed in models with these general characteristics may \textit{not} display the secondary peaks corresponding to quasilinear combinations of the fundamental $l=0$ and $l=2$ modes. The absence of these features could be actually a smoking gun for the presence of additional scalar degrees of freedom in our universe.

On the other hand, for the other model studied in this work (MO100), coupling between fluid and scalar field oscillations is weak. As a consequence, energy is drained more slowly by the scalar field, and all modes have a decay timescale much larger than the dynamical timescale of a NS. A peculiar feature of this model is that the modified (fluid-led) $F$-mode is still the fundamental mode of the NS. The fact that we are able to see peaks corresponding to the nonlinear self-coupling of this mode in our simulations (cf.~Figs.~\ref{fig:S_MO_DFT}, \ref{fig:M_MO_dft}, and \ref{fig:MS_MO_dft}) suggests that it should continue to couple strongly with the $l=2$ mode, giving rise to similar phenomenology as in GR. Moreover, the slightly different frequency with respect to GR could easily be mimicked by a modification to the nuclear EOS. 

Note, however, that weak coupling not necessarily means that the spectrum is just slightly altered with respect to GR, as families of scalar-led modes may lead to qualitatively new features. In the MO100 model, for a certain range of central densities, we see what could be thought of as a splitting of the GR $H_1$ mode, as the fundamental scalar-led mode resides in a similar frequency range---see Fig.~\ref{fig:linearspectrumMO}. In some situations, these overtones could be excited (cf. Fig.~\ref{fig:S_MO_DFT}), giving rise to a very different pattern with respect to GR. This highlights the importance of conducting systematic studies of linear perturbations in relevant alternative theories of gravity. 

Finally, it is worthwhile to mention that next-generation GW detectors may also have enough sensitivity to detect the imprints of NS oscillations during the inspiral phase. Modes excited during this phase can drain energy from the orbital motion, resulting in a different phase evolution with respect to GR. We leave a detailed investigation of this scenario to future work. 

\section*{Acknowledgments}

We are grateful to the organizers of the Workshop on Black Holes and Neutron Stars in Modified Gravity, Meudon (18-20 November 2019) where the discussions that led to the present work were initiated. N.S. acknowledges support by the COST actions CA16214 ``PHAROS", CA16104 ``GWVerse" and CA18108 ``QG-MM". N.S. gratefully acknowledges the Italian Instituto Nazionale di Fisica Nucleare (INFN), the French Centre National de la Recherche Scientifique (CNRS) and the Netherlands Organization for Scientific Research, for the construction and operation of the Virgo detector and the creation and support of the EGO consortium. N.O. acknowledges financial support by the CONACyT grants ``Ciencia de Frontera" 140630 and 376127, and by the UNAM-PAPIIT grant IA100721. R.M. acknowledges partial funding from the National Council
for Scientific and Technological Development (CNPq) and by the Carlos Chagas Filho Research Support Foundation (FAPERJ).

\bibliography{lib_qnm_stt}

\end{document}